\documentclass [a4paper,11pt]{article}
\usepackage[english]{babel}
\usepackage{graphicx,amsmath,amsfonts,amssymb,slashed,caption,amscd,cancel, tensor,color,booktabs}
\usepackage{hyperref}
\definecolor{MyBlue}{rgb}{0.15,0.15,0.70}

\hypersetup{
colorlinks=true,
citecolor=MyBlue,
linkcolor=MyBlue,
urlcolor=MyBlue
}
\newcommand{\be}{\begin{equation}}
\newcommand{\ee}{\end{equation}}
\newcommand{\beq}{\begin{equation}}
\newcommand{\eeq}{\end{equation}}
\newcommand{\bea}{\begin{eqnarray}}
\newcommand{\eea}{\end{eqnarray}}

\newcommand{\R}{R}
\newcommand{\G}{G}

\newcommand{\ku}{(K_1)}
\newcommand{\kd}{(K_2)}
\newcommand{\nd}{{\dot n}}

\newcommand{\M}{{\cal M}}
\newcommand{\Y}{Y}

\newcommand{\SK}{{\cal S}}
\newcommand{\YY}{{\cal Y}}

\newcommand{\pid}{{\dot \pi}}

\newcommand{\gpi}{{\nabla\pi}}
\usepackage[stable]{footmisc}
\def\d{\delta}

\def\Mp{M_{\rm Pl}}
\def\MM{M_{*}}
\newcommand{\mfs}{{m}_4^2}
\newcommand{\tmfs}{{\tilde m}_4^2}
\def\s{\sigma}
\def\ga{\Gamma}
\def\Rb{\bar R}

\def\dkmu2{\delta K_{\mu \nu}\delta K^{\mu \nu}}
\def\pmu2{  \phi_{\mu \nu}\phi^{\mu \nu}}

\newcommand{\gammah}{h}
\newcommand{\Atwo}{{A_2}}
\newcommand{\Athree}{{A_3}}
\newcommand{\Afour}{{B_4}}
\newcommand{\Afive}{{B_5}}
\newcommand{\Bfour}{{A_4}}
\newcommand{\Bfive}{{A_5}}

\begin{document}

\begin{center}
\Large{\textbf{Exploring gravitational theories beyond Horndeski}} \\[1cm] 
 
\large{J\'er\^ome Gleyzes$^{\rm a,b}$,  David Langlois$^{\rm c}$, Federico Piazza$^{\rm c,d,e}$ and Filippo Vernizzi$^{\rm a}$}
\\[0.5cm]

\small{
\textit{$^{\rm a}$ CEA, IPhT, 91191 Gif-sur-Yvette c\'edex, France \\ CNRS,  URA-2306, 91191 Gif-sur-Yvette c\'edex, France}}

\vspace{.2cm}

\small{
\textit{$^{\rm b}$ Universit\'e Paris Sud, 15 rue George Cl\'emenceau, 91405,  Orsay, France}}

\vspace{.2cm}

\small{
\textit{$^{\rm c}$  APC, (CNRS-Universit\'e Paris 7), 10 rue Alice Domon et L\'eonie Duquet, 75205 Paris, France \\ 
}}
\vspace{.2cm}

\small{
\textit{$^{\rm d}$ Aix Marseille Universit\'e, CNRS, CPT, UMR 7332, 13288 Marseille,  France. \\
}}
\vspace{.2cm}

\small{
\textit{$^{\rm e}$ Physics Department and Institute for Strings, Cosmology, and Astroparticle Physics,\\
  Columbia University, New York, NY 10027, USA}}

\vspace{0.5cm}
\today

\end{center}

\vspace{2cm}

\begin{abstract}
We have recently proposed a new class of gravitational scalar-tensor theories free from  Ostrogradski instabilities, in Ref.~\cite{Gleyzes:2014dya}. 
As they generalize  Horndeski theories, or ``generalized'' galileons,  we call them G$^3$. 
These theories possess a  simple  formulation when the time hypersurfaces are chosen to coincide with the uniform scalar field hypersurfaces. We confirm that they contain only three propagating degrees of freedom by presenting the details of the Hamiltonian formulation. 
We examine the coupling between these theories and matter. Moreover, we investigate how they transform under a disformal redefinition of the metric.
Remarkably, these theories are preserved by disformal transformations that depend on the scalar field gradient, which also allow to map subfamilies of G$^3$ into Horndeski theories.
\end{abstract}

\newpage 
\tableofcontents

\vspace{.5cm}

\section{Introduction}
The fact that current cosmological observations consistently point to a recent phase of  accelerated expansion has boosted the exploration of  alternative theories of gravity (see \emph{e.g.}~\cite{Clifton:2011jh} for a review), that could provide a more natural explanation than simply a cosmological constant.
 Even if these efforts have not led to a compelling or even realistic model, these research activities
 have deepened  our understanding of gravity  by highlighting the theoretical and observational constraints that alternatives to general relativity must satisfy.

Many models of modified gravity involve the presence of at least one scalar degree of freedom  in addition to the two tensor degrees of freedom of general relativity. The underlying scalar field can sometimes be hidden in the explicit formulation of the theory. A typical example is $f(R)$ theory, where the Lagrangian is written as a function of the Ricci scalar $R$, but which can be reformulated as a manifestly scalar-tensor theory (see \emph{e.g.}~\cite{Chiba:2003ir}).

A minimal requirement on alternative theories is the absence of ghost-like instabilities within their domains of validity (see {\em e.g.}~\cite{Burgess:2014lwa} on this point).  According to the so-called Ostrogradski's theorem, such instabilities arise in theories characterized by  a non-degenerate Lagrangian\footnote{A Lagrangian $L(q,\dot q, \ddot q)$ is said to be nondegenate if $\partial^2 L/\partial \ddot q^2\neq 0$} with higher time derivatives (see {\em e.g.}~\cite{woodard}). The simplest example is the Lagrangian
\beq 
L=\frac12 \ddot q^2,
\eeq
which leads to fourth-order equations of motion. In the Hamiltonian formulation, an extra degree of freedom appears so that the corresponding phase space  is four-dimensional, with a Hamiltonian that depends linearly on one of the momenta and is thus (kinetically) unbounded from below. In this case the extra degree of freedom is  a ghost and the theory is not viable.

Not all theories containing higher-order time derivatives in the Lagrangian suffer from Ostrogradski instabilities. In particular, this is the case 
for theories that lead to second-order equations of motion, such as the much studied galileon models~\cite{NRT}, briefly reviewed in Sec.~\ref{gt}. 
Although originally introduced  in Minkowski, the galileon Lagrangians can be extended  to general curved spacetimes by promoting the derivatives to {\em covariant} derivatives. However, as discussed in Sec.~\ref{cg}, maintaining second-order equations of motion with respect to spacetime derivatives requires the addition  of suitable gravitational ``counterterms" \cite{Deffayet:2009wt,Deffayet:2009mn}. The largest  class of these Generalized Galileons~\cite{Deffayet:2011gz}, or G$^2$, turns out to be equivalent  to the more ancient Horndeski's theories \cite{horndeski}, which correspond to  the most general scalar-tensor theories with second-order field equations.

Although Horndeski theories are often considered as the most general scalar-tensor theories immune from Ostrogradski's instabilities, we have recently showed that this is not the case and proposed a new class of scalar-tensor theories, reviewed in Sec.~\ref{G3} (see also Appendix \ref{Covariant1} for the details of the calculations), that do not suffer from such instabilities \cite{Gleyzes:2014dya}. Since our theory contains generalized galileons (Horndeski) as a special limit, we dubbed it ``Generalized Generalized Galileons" or G$^3$ for brevity. It turns out that our theories have the same decoupling limit as Horndeski theories, as briefly showed at the end of Sec.~\ref{G3}.

The stability properties of G$^3$ are most easily seen by using the ADM formalism applied to the uniform scalar field hypersurfaces (also called unitary gauge formulation). In this formulation, the scalar field does not appear explicitly as it is part of the degrees of freedom of the metric, and the action depends only on first time derivatives of the metric (the ``velocities"), as generally expected from healthy theories. Indeed, the Hamiltonian analysis confirms the absence of unwanted extra degrees of freedom, and thus the absence of Ostrogradski instabilies~\cite{Gleyzes:2014dya}. In Sec.~\ref{H} of the present article we give more details about the derivation of the Hamiltonian  and about the counting of the degrees of freedom, which depends on the number and nature (first or second class) of the constraints between canonical variables. Our analysis clearly proves that our theories contain only three degrees of freedom and do not suffer from Ostrogradski instabilities, as stated in \cite{Gleyzes:2014dya}.

Hints that one could go beyond Horndeski theories without encountering fatal instabilities appeared in our previous work  \cite{GLPV}, where we studied the most general quadratic Lagrangian for linear perturbations about a homogenous and isotropic spacetime that does not induce higher derivatives on the linear propagating scalar degree of freedom. Such a Lagrangian contains an additional term, which is absent in Horndeski theories. In Sec.~\ref{M1} we review this analysis  of linear perturbations and we extend it in Sec.~\ref{M2} by including some matter field, detailing the analysis of \cite{Gleyzes:2014dya}. For convenience, we describe matter by means of a scalar field with non-standard kinetic term, which can be formulated in terms of a simple Lagrangian and which is characterized by a nontrivial speed of sound. We are thus able to derive a quadratic Lagrangian that includes both metric and matter perturbations in the unitary gauge. A similar calculation was presented in \cite{Gergely:2014rna}, and generalized to several matter scalar fields in \cite{Kase:2014yya}. We also give an equivalent treatment for perfect fluid matter by working directly with the equations of motion written in the Newtonian gauge, in Sec.~\ref{HEsect}. For this analysis we find it convenient to employ  the notation proposed in Ref.~\cite{Sawicki},  based on the effective approach to cosmological perturbations for dark energy, introduced in \cite{EFTOr,Bloomfield:2012ff,GLPV,Bloomfield:2013efa}. In Appendix \ref{app:BB} we review the connection between the different notations employed in these references. 
 
Other fissures in the standard lore concerning Horndeski theories were pointed out in \cite{Zumalacarregui:2013pma}, which studied scalar-tensor theories generated by  disformal relations 
\cite{Bekenstein:1992pj}
\beq
\tilde g_{\mu\nu}=\Omega^2(X,\phi) g_{\mu\nu}+\ga(X,\phi) \partial_{\mu} \phi \partial_{\nu} \phi \,,
\eeq
where $X\equiv g^{\mu\nu} \partial_{\mu} \phi \partial_{\nu} \phi$. In particular, it was shown that starting from an action consisting of the Einstein-Hilbert term for $\tilde g_{\mu\nu}$ and of  a standard action for $\phi$, one obtains  equations of motion for $g_{\mu\nu}$ and $\phi$ that are higher order but can be combined so that the dynamics is only second order. This is another example beyond Horndeski that is not Ostrogradski unstable. Interestingly, a very similar argument has been invoked in the context of ghost-free massive gravity in \cite{deRham:2011qq}.

It is  natural to wonder whether our theories could be formulated in a similar way, i.e.~derived via a disformal transformation  from a theory belonging to the Horndeski class. We discuss this issue in Sec.~\ref{D} and find that our general theory cannot be 
derived from Horndeski via a disformal transformation. Remarkably however,  the two non-Horndeski  pieces contained in our Lagrangian can be \emph{separately}  derived from a Horndeski Lagrangian combined with a disformal transformation. 
Since the disformal transformation that we consider conserves the number of degrees of freedom, this proves that our two non-Horndeski pieces are separately equivalent to a subset of Horndeski theories. In Appendix \ref{app:CC} we explicitly check in Newtonian gauge that the disformal metric redefinition de-mixes part of the kinetic couplings (the part  containing higher derivatives) between the scalar field and the metric. In this respect, the disformal transformations considered here are analogous to the field redefinition removing higher derivatives discussed in the context of massive gravity in \cite{deRham:2011qq}.
Since the two disformal transformations are distinct for the two non-Horndeski pieces of G$^3$, the procedure cannot be applied to the whole Lagrangian. However, the fact that these pieces can be mapped to Horndeski provides an alternative way to show the  healthy behavior of our theories.
Using a disformal transformation, in Sec.~\ref{sec-6:5} we provide an example of naively higher-derivative equations of motion which can be reduced to second order ones, generalizing the treatment of~\cite{Zumalacarregui:2013pma}.

%%%%%%%%%%%%%%%%%%%%%%%%%%%%%%%%%%%%%%%%%%%%%%%%%%%%%%%%%%%%%%%%%%%%%%%%%
%%%%%%%%%%%%%%%%%%%%%%%%%%%%%%%%%%%%%%%%%%%%%%%%%%%%%%%%%%%%%%%%%%%%%%%%%
\section{Galileons and  Horndeski theories}
%%%%%%%%%%%%%%%%%%%%%%%%%%%%%%%%%%%%%%%%%%%%%%%%%%%%%%%%%%%%%%%%%%%%%%%%%
%%%%%%%%%%%%%%%%%%%%%%%%%%%%%%%%%%%%%%%%%%%%%%%%%%%%%%%%%%%%%%%%%%%%%%%%%
\newcommand{\Gtwo}{G_2{}}
\newcommand{\Gthree}{G_3{}}
\newcommand{\Gfour}{G_4{}}
\newcommand{\Gfive}{G_5{}}
\newcommand{\Ftwo}{F_2{}}
\newcommand{\Fthree}{F_3{}}
\newcommand{\Ffour}{F_4{}}
\newcommand{\Ffive}{F_5{}}

\subsection{Galileon theories}
\label{gt}

 One of  the most explored frameworks for infra-red modifications of gravity is the so-called galileon theory~\cite{NRT}, which distills and generalizes the interesting features of the DGP scenario~\cite{DGP} and emerges in the decoupling limit of massive gravity~\cite{deRham:2010ik}. 

Galileon theories can be seen as the effective theory of a Goldstone boson $\phi$ in Minkowski space, that is invariant under a generalized shift symmetry, 
\begin{equation} \label{symmetry}
\phi(x) \ \rightarrow \ \phi(x) + b_\mu x^\mu + c,
\end{equation}
for the five arbitrary parameters $b_\mu$ and $c$.  Only in Minkowski can we arbitrarily choose a \emph{constant} vector field $b^\mu$ and thus this is where galileon theories are naturally set.   
At lowest order in derivatives, there exists a limited number of Lagrangian terms invariant under~\eqref{symmetry}, with schematic form ${\cal L}_n \sim (\partial \phi)^2 (\partial^2 \phi)^{n-2}$, where $n\leq 5$ in four dimensions. Such operators are protected by the symmetry~\eqref{symmetry} against quantum corrections~\cite{Luty:2003vm,Nicolis:2004qq}.

These theories can be most naturally formulated as~\cite{NRT}
\begin{equation}
{\cal L}_{n+1}^{\rm gal,1} = \left({\cal A}^{\mu_1 \dots \mu_n \nu_1 \dots \nu_n} \phi_{\mu_1} \phi_{\nu_1}\right) \phi_{\mu_2 \nu_2} \dots  \phi_{\mu_n \nu_n} \, ,
\end{equation}
where ${\cal A}^{\mu_1 \dots \mu_n \nu_1 \dots \nu_n}$ is a tensor separately antisymmetric 
in the indices $\mu$'s and  $\nu$'s and symmetric under the exchange $\{\mu_i\} \leftrightarrow \{\nu_i\}$, \emph{e.g.} $\mathcal{A}^{\mu_1 \mu_2 \nu_1 \nu_2}\propto g^{\mu_1\nu_1}g^{\mu_2\nu_2}-g^{\mu_1\nu_2}g^{\mu_2\nu_1}$ (see {\em e.g.}~the nice review~\cite{Deffayet:2013lga} for technical details). In the above expression and in the rest of this section, we use the shorthand notation $\phi_\mu\equiv\nabla_\mu \phi$, $\phi_{\mu\nu}\equiv \nabla_\nu\nabla_\mu\phi$ for convenience.
More explicitly, the galileon Lagrangians are written as linear combinations of the five following Lagrangians: 
\begin{align}
 L_2^{\rm gal,1}  =&\ X \;, \label{L22g}\\ 
 L_3^{\rm gal,1}  =&\  X \Box \phi -  \phi_{\mu}  \phi^{\mu \nu}  \phi_\nu \;, \label{L33g} \\
L_4^{\rm gal,1} =&\ X \big[(\Box \phi)^2 - \phi_{\mu \nu} \phi^{\mu \nu}\big]  - 2 (\phi^{\mu} \phi^{\nu} \phi_{\mu \nu} \Box \phi - \phi^{\mu}  \phi_{\mu \nu} \phi_{\lambda} \phi^{\lambda \nu})\; ,  \label{L44g} \\
L_5^{\rm gal,1}  =&\ X \big[ (\Box \phi)^3 - 3 (\Box \phi) \phi_{\mu \nu} \phi^{\mu \nu} +2 \phi_{\mu \nu} \phi^{\nu \rho} \phi^\mu_{\ \rho}\big] \label{L55g} \\
&-  3 \big[ (\Box \phi)^2 \phi_\mu \phi^{\mu \nu} \phi_\nu - 2 \Box \phi \phi_\mu \phi^{\mu \nu} \phi_{\nu \rho} \phi^\rho  - \phi_{\mu \nu} \phi^{\mu \nu} \phi_\rho \phi^{\rho \lambda} \phi_\lambda+ 2 \phi_\mu \phi^{\mu \nu} \phi_{\nu \rho} \phi^{\rho \lambda} \phi_\lambda \big]\; . \nonumber 
\end{align}

In flat space there exist alternative (in fact, infinite) versions of galileon Lagrangians, equivalent up to total derivatives. A particularly compact and popular choice (called ``form 3" in~\cite{Deffayet:2013lga}) is
\begin{align}
 L_2^{\rm gal,3}  =&\ X \;, \label{L22g3}\\ 
 L_3^{\rm gal,3}  =&\  \frac32 X \Box \phi  \;, \label{L33g3} \\
L_4^{\rm gal,3} =&\ 2 X \big[(\Box \phi)^2 - \phi_{\mu \nu} \phi^{\mu \nu}\big]\; ,  \label{L44g3} \\
L_5^{\rm gal,3}  =&\ \frac52 X \big[ (\Box \phi)^3 - 3 (\Box \phi) \phi_{\mu \nu} \phi^{\mu \nu} +2 \phi_{\mu \nu} \phi^{\nu \rho} \phi^\mu_{\ \rho}\big] \label{L55g3}\; , 
\end{align}
where we have chosen the normalization factors in order to be consistent with the original expressions~\eqref{L22g}-\eqref{L55g}.

\subsection{Coupling to gravity and Horndeski theories}
\label{cg}

By going from~\eqref{L22g}-\eqref{L55g} to~\eqref{L22g3}-\eqref{L55g3} we have exchanged the order of partial derivatives, which can be consistently done in flat space. But in general curved spaces, while doing so for $L_4$ and $L_5$ we have to pay a commutator proportional to the curvature. Indeed, by taking $f$ as a general function of $X$, we find that the two main blocks of terms appearing in $L_4^{\rm gal,1}$ and $L_5^{\rm gal,1}$ are related by, respectively,
\be
f \big[(\Box \phi)^2 - \phi_{\mu \nu} \phi^{\mu \nu}\big]  = - 2 f_X (\phi^{\mu} \phi^{\nu} \phi_{\mu \nu} \Box \phi - \phi^{\mu}  \phi_{\mu \nu} \phi_{\lambda} \phi^{\lambda \nu}) + f \ {}^{(4)} \!R^{\mu \nu} \phi_\mu \phi_\nu  + {\rm boundary \ terms}\, , 
\ee
and
\be
\begin{split}
 f \big[ (\Box \phi)^3 &- 3 (\Box \phi) \phi_{\mu \nu} \phi^{\mu \nu} +2 \phi_{\mu \nu} \phi^{\nu \rho} \phi^\mu_{\ \rho}\big] =  \\ 
&- 2 f_X \big[ (\Box \phi)^2 \phi_\mu \phi^{\mu \nu} \phi_\nu - 2 \Box \phi \phi_\mu \phi^{\mu \nu} \phi_{\nu \rho} \phi^\rho  - \phi_{\mu \nu} \phi^{\mu \nu} \phi_\rho \phi^{\rho \lambda} \phi_\lambda+ 2 \phi_\mu \phi^{\mu \nu} \phi_{\nu \rho} \phi^{\rho \lambda} \phi_\lambda \big] \\
&  - 2 X f \, \left({}^{(4)} \!R_{ \mu \sigma \rho \nu}\phi^\mu \phi^{\rho \sigma} \phi^\nu + {}^{(4)} \!R_{\mu \nu} \phi_\sigma \phi^{\mu \sigma} \phi^\nu - {}^{(4)} \!R_{\mu \nu} \phi^\mu \phi^\nu \Box \phi \right) + {\rm boundary \ terms} \, .
\end{split}
\ee
This also means that the different versions of the galileon Lagrangians, which are all equivalent in flat space, correspond to genuinely different theories once minimally coupled to gravity by trading ordinary derivatives for covariant derivatives. 
Of course, as realized in~\cite{Deffayet:2009wt}, the minimally coupled versions of galileons $L_4$ and $L_5$ bring higher (third order) derivatives into the equations of motion. For example, by varying $X (\Box \phi)^{2}$ with respect to $\phi$, one ends up with terms containing two derivatives hitting on a Christoffel symbol, i.e., three derivatives of the metric.  In order to get rid of such higher derivatives, the authors of~\cite{Deffayet:2009wt} added to $L_4^{\rm gal,1}$ and $ L_5^{\rm gal,1}$  suitable gravitational ``counterterms" and thus ``re-discovered" Horndeski theories~\cite{horndeski}, which can be described by an arbitrary linear combination of the Lagrangians
\begin{align}
L_2^{H} [G_2] \equiv & \; \Gtwo(\phi,X)\;,  \label{L2} \\
L_3^{H} [G_3] \equiv & \; \Gthree(\phi, X) \, \Box \phi \;, \label{L3} \\
L_4^{H} [G_4] \equiv &\;\Gfour(\phi,X) \, {}^{(4)}\!R - 2 \Gfour_{X}(\phi,X) (\Box \phi^2 - \phi^{ \mu \nu} \phi_{ \mu \nu}) \;, \label{L4} \\
L_5^{H} [G_5] \equiv & \;\Gfive(\phi,X) \, {}^{(4)}\!G_{\mu \nu} \phi^{\mu \nu} +\frac13  \Gfive_{X} (\phi,X)  (\Box \phi^3 - 3 \, \Box \phi \, \phi_{\mu \nu}\phi^{\mu \nu} + 2 \, \phi_{\mu \nu}  \phi^{\mu \sigma} \phi^{\nu}_{\  \sigma}) \;,\label{L5}
\end{align}
following the presentation given in Ref.~\cite{Deffayet:2011gz}.

\section{Beyond Horndeski: G$^3$}
\label{G3}
As we have recently shown  in \cite{Gleyzes:2014dya}, it turns out that it is possible to extend the Horndeski Lagrangians presented above without encountering ghost-like Ostrogradski instabilities. In order to introduce these theories, it is much simpler to use the so-called unitary gauge, where  the uniform scalar field ($\phi=$ const)  hypersurfaces coincide with  constant-time hypersurfaces. To do so, we assume that the gradient of the scalar field, $\partial_\mu\phi$, is time-like. 
Using an ADM decomposition of the metric,
\be
ds^2=- N^2 dt^2  +\gammah_{ij} (dx^i+ N^i dt)(dx^j+ N^j dt ) \;,
\ee
we write the Lagrangian density  in terms of the  intrinsic and extrinsic 3-d curvature tensors of the spatial slices, respectively denoted $R_{ij}$ and $K_{ij}$, their traces, $\R\equiv h^{ij}\R_{ij}$, $K\equiv h^{ij}K_{ij}$,  as well as the lapse function $N$.
The  theories presented in \cite{Gleyzes:2014dya} are then given by the action
\be
S= \int d^4 x \sqrt{-g} ( L_2 +L_3 +L_4+L_5)\;, \label{full_action}
\ee
with
\be
\label{Lagrangian_uni}
\begin{split}
 L_2 & \equiv \Atwo(t,N)\;, \\ 
 L_3 &\equiv\Athree (t,N) K\;, \\
L_4 &\equiv \Bfour (t,N) \big(K^2 - K_{ij}K^{ij} \big) + \Afour (t,N) \R\;, \\
L_5 &\equiv \Bfive(t,N) \big(K^3 - 3 K K_{i j}K^{i j} + 2  K_{i j}  K^{i k} K^j_{\ k} \big)  + \Afive (t, N) K^{ij} \bigg( \R_{ij} - \frac12 \gammah_{ij} R \bigg)   \;,
\end{split}
\ee
where $A_a$ and $B_a$ ($a=2,3,4,5$) are generic functions of $t$ and $N$. Let us remind that, in terms of ADM variables, the extrinsic curvature reads 
\be
\label{extrinsic}
K_{ij} = \frac{1}{2N} \big(\dot h_{ij} - D_i N_j - D_j N_i \big) \;, 
\ee
where $D_i$ is the spatial covariant derivative. The combination $K^2 - K_{ij}K^{ij}$ in the third line is the usual GR kinetic term.
Indeed,  when $\Afour=-\Bfour=1/(16 \pi G)$,  while the other coefficients vanish, the above action corresponds to the Einstein-Hilbert action up to boundary terms, as can  be easily seen upon using
 the Gauss-Codazzi relation (see eq.~\eqref{GC1} in App.~\ref{Covariant1}). 
In this case the action becomes fully 4-d diff invariant and there are no propagating scalar degrees of freedom.

We now rewrite the above Lagrangians in a manifestly covariant form, \emph{i.e.}~in terms of $\phi$ and its spacetime derivatives. The dependence on $t$ and $N$ of the functions $A_a$ and $B_a$ will turn into a dependence on  $\phi$ and $X \equiv g^{\mu \nu} \partial_\mu \phi \partial_\nu \phi$,  since  $\phi = \phi_0(t)$  and $X= -\dot\phi_0^2(t)/N^2$ in our ADM formulation. We can then introduce the unit vector normal to the uniform $\phi$ hypersurfaces,
\be
\label{defN}
n_\mu \equiv - \frac{\partial_{\mu} \phi}{\sqrt{-X}}\, ,
\ee
and define the extrinsic curvature as
\be
\label{defK}
K_{\mu \nu} \equiv (g^\sigma_{\ \mu} + n^\sigma n_\mu ) \nabla_\sigma n_\nu \, .
\ee
Using this expression and $K\equiv \nabla_\mu n^\mu$, and denoting the derivation by a lower index, {\em e.g.}~$A_{2X} \equiv \partial A_2 /\partial X$, the above Lagrangians can be rewritten, after lengthy but straightforward manipulations explicitly given in App.~\ref{Covariant1}, as \cite{Gleyzes:2014dya}
\begin{align}
 L_2 & = L^H_2[A_2] \;, \label{L22}\\ 
 L_3 & =  L^H_3[C_3+2 X C_{3X}] + L^H_2[XC_{3\phi}] \;, \label{L33} \\
L_4 &= L_4^H[B_4]  + L^H_3[C_4+2 X C_{4X}] + L^H_2[XC_{4\phi}]  -  \frac{\Afour+\Bfour - 2 X \Afour_{X}}{X^2} L_4^{\rm gal,1} \;, \label{L44} \\
L_5 & =L_5^H[G_5]+ L_4^H[C_5] + L_3^H[D_5 + 2 X D_{5X}]  + L_2^H[X D_{5 \phi}] +\frac{X \Afive_X + 3  \Bfive}{3 (-X)^{5/2}} L_5^{\rm gal,1}    \label{L55}  \, ,
\end{align}
where $A_a$ and $B_a$ are now functions of $\phi$ and $X$, $A_a = A_a(\phi,X)$, $B_a = B_a(\phi,X)$,   and $C_3$, $C_4$, $C_5$, $D_5$ and $G_5$ are defined as
\be
\begin{split}
C_3 & \equiv \frac12  \int \Athree (-X)^{-3/2}     \, dX \;, \\
C_4 &\equiv - \int  \Afour_{\phi} (-X)^{-1/2}\, dX \;, \\
C_5 & \equiv - \frac14 X \int B_{5 \phi} (-X)^{-3/2} d X \;, \\
D_5 & \equiv -\int  C_{5\phi} (-X)^{-1/2}\, dX \;, \\
 G_{5} & \equiv - \int  \Afive_{X}(-X)^{-1/2} \,dX \;.
\end{split}
\ee
If  
 $\Bfour$ and $\Bfive$ are related to $\Afour$ and $\Afive$ by
\be
\Bfour= - \Afour + 2X \Afour_{X}\;, \qquad \Bfive = - X \Afive_{X}/3\;, \label{gal}
\ee
the last terms of both eqs.~\eqref{L44} and~\eqref{L55} vanish. In this case one is left only with the Horndeski Lagrangians, which manifestly shows that eqs.~\eqref{L22}--\eqref{L55} (and thus action~\eqref{full_action})  contain Horndeski theories.
In general, the functions  $\Bfour$ and $\Bfive$ are completely free, which means that our theories contain two additional free functions with respect to the Horndeski ones.

It is straighforward to see that the minimally coupled versions of the original galileons proposed in~\cite{NRT}, \eqref{L22g}--\eqref{L55g}, are contained in eqs.~\eqref{L22}--\eqref{L55} by the choice of functions  $\Afour=0$, $\Afive=0$, $A_2 = X$, $A_3=3X/2$, $A_4=-X^2$ and $A_5=(-X)^{5/2}$. As a corollary, $L_{4}^{\rm gal,1}$ and $L_{5}^{\rm gal,1}$ are already healthy without the need of additional gravitational counterterms.  In other words,
the straightforward covariantization of galileons, i.e.~substituting ordinary derivatives with covariant derivatives, is a viable covariantization. It should be noted, however, that galileon symmetry remains broken by terms proportional to the curvature, regardless of the chosen covariantization procedure.

Finally, before concluding this section, let us briefly comment on the decoupling limit of eqs.~\eqref{L22}--\eqref{L55}. In Ref.~\cite{KNT},  the decoupling limit of Horndeski theories has been studied by expanding the metric $g_{\mu \nu}$ around Minkowski and the scalar field $\phi$ around a constant background value. In doing so, the following scaling of the functions $G_a(\phi, X)$ introduced in eqs.~\eqref{L2}--\eqref{L5} was assumed \cite{Narikawa:2013pjr},
\be
G_2 \sim \Lambda_3^3 \Mp \;, \quad G_3 \sim \Mp \;, \quad G_4 \sim \Mp^2 \;, \quad G_5 \sim \Lambda_3^{-3} \Mp^2 \;,
\ee
where $\Lambda_3$ is a mass scale 
which may be associated to the current accelerated expansion of the universe (in which case $\Lambda^3_3 \sim \Mp H_0^2$) and $\Mp$ is the Planck mass. The decoupling limit is defined as  $\Mp \to \infty$ while  $\Lambda_3$ remains constant. It is easy to see that taking this limit in eqs.~\eqref{L22}--\eqref{L55}  leads to the same decoupling limit found in \cite{KNT} for Horndeski, but with different dimensionless parameters. 
This is clearly the case for eqs.~\eqref{L22} and \eqref{L33}, because they are equivalent to the Horndeski Lagrangians $L_2^H$ and $L_3^H$. Equations~\eqref{L44} and \eqref{L55} contain non-Horndeski pieces, respectively $L_4^{\rm gal,1}$ and $L_5^{\rm gal,1}$. By expanding these terms in scalar field and metric perturbations, the only contributions that do not vanish in the decoupling limit are galileons, i.e.,
\be
-   \frac{\Afour+\Bfour - 2 X \Afour_{X}}{X^2}  L_4^{\rm gal,1} \sim \Lambda_3^{-6}L_4^{\rm gal,1} \;, \qquad \frac{X \Afive_X + 3  \Bfive}{3 (-X)^{5/2}} L_5^{\rm gal,1} \sim \Lambda_3^{-9} L_5^{\rm gal,1}\;,
\ee
where the  functions $({\Afour+\Bfour - 2 X \Afour_{X}})/{X^2}$ and $({X \Afive_X /3+   \Bfive})/(-X)^{5/2}$  are evaluated on the background. In conclusion, operators leading to higher-derivative equations of motion in eqs.~\eqref{L44} and \eqref{L55} are also higher order in the decoupling limit.

%%%%%%%%%%%%%%%%%%%%%%%%%%%%%%%%%%%%%%%%%%%%%%%%%%%%%%%%%%%%%%%%%%%%%%%%%
%%%%%%%%%%%%%%%%%%%%%%%%%%%%%%%%%%%%%%%%%%%%%%%%%%%%%%%%%%%%%%%%%%%%%%%%%
\section{Hamiltonian analysis}
\label{H}
%%%%%%%%%%%%%%%%%%%%%%%%%%%%%%%%%%%%%%%%%%%%%%%%%%%%%%%%%%%%%%%%%%%%%%%%%
%%%%%%%%%%%%%%%%%%%%%%%%%%%%%%%%%%%%%%%%%%%%%%%%%%%%%%%%%%%%%%%%%%%%%%%%%
\def\tH{{\tilde H}}
\def\M{{\cal M}}
\def\T{{\cal T}}
\def\H{{\cal H}}

As discussed in the introduction, theories that contain higher-order time derivatives often lead to lethal Ostrogradski instabilities. The presence of  higher derivatives manifests itself in the form of extra degrees of freedom that behave like ghosts (i.e.~negative energy states). For instance, the dynamics of  a system with a nondegenerate Lagrangian of the form $L(q, \dot q, \ddot q)$ is described by a 4-dimensional phase space, corresponding to two degrees of freedom, one of which behaves like a ghost (see {\em e.g.}~\cite{woodard}).

In the ADM formulation, our Lagrangian (\ref{Lagrangian_uni}) depends on the dynamical quantities $h_{ij}$ and their ``velocities'' $K_{ij}$: in this sense, it is already evident that the Lagrangian does not contain higher-order time derivatives and that Ostrogradski instabilities should not be there. In order to confirm this intuition, we now perform the Hamiltonian analysis for the  Lagrangian (\ref{Lagrangian_uni}) and show that the number of degrees of freedom remains three---i.e.~two tensor modes and one scalar mode, thus excluding the appearance of dangerous extra degrees of freedom. The present analysis details that of  \cite{Gleyzes:2014dya} and confirms its conclusions.

The phase space of our theory is described by the variables $h_{ij}$, $N$, $N^i$ and their  conjugate momenta, given respectively by 
\be
\label{invert}
\begin{split}
\pi^{ij}\equiv\frac{\partial {\cal L}}{\partial \dot\gammah_{ij}}=
\frac{\sqrt{h}}{2 }
&
\left[
\left(A_3+2A_4 K+3 A_5(K^2-K_{lm}K^{lm})\right) h^{ij}
\right.
\cr
&
\left.
-2(A_4+3A_5 K)K^{ij} +6 A_5 K^i_l K^{lj}+B_5\left(R^{ij}-\frac12 R h^{ij}\right)
\right]\,,
\end{split}
\ee
and 
\be 
\label{primary}
\pi_N \equiv \frac{\partial {\cal L}}{\partial \dot N}=0\, \qquad \pi_i\equiv \frac{\partial {\cal L}}{\partial \dot N^i}=0\,.
\ee
The absence of time derivatives of the lapse $N$ and the shift $N^i$ in the action implies that their conjugate momenta automatically vanish. The relations $\pi_N=0$ and $\pi_i=0$ can thus be seen as restrictions of the initial 20-dimensional phase space, corresponding to so-called {\it primary constraints}.
So far, the situation  is quite similar to that of  pure  general relativity.

The canonical Hamiltonian is then obtained  via the Legendre transform of the Lagrangian,
\be
\label{Hamiltonian}
H\equiv \int d^3\vec  x \left[\pi^{ij} \dot\gammah_{ij} -{\cal L}\right]\,.
\ee
The Hamiltonian is expressed in terms of the canonical variables, which means that, in principle, one must invert the relation in (\ref{invert})  to obtain $\dot\gammah_{ij}$ as a function of $\pi^{ij}$.  Because of the presence of primary constraints, the time evolution is governed by the extended Hamiltonian,
\be
\tH=H+\int d^3\vec x \left[\lambda_N\, \pi_N+\lambda^i\,\pi_i\right]\,,
\ee
where $\lambda_N$ and $\lambda_i$ play the role of Lagrange multipliers. For any function $F$ defined on the phase space, its time evolution is given by
\be
\frac{d}{dt}F=\frac{\partial F}{\partial t}+\big\{F, \tH\big\}\,.
\ee
The Poisson bracket in the above formula is defined, as usual, by the expression
\be
\left\{F,G\right\}\equiv \sum_A\int d^3 \vec  x\left(\frac{\delta F}{\delta \phi^A( \vec  x)}\frac{\delta G}{\delta \pi_A(\vec x)}-\frac{\delta F}{\delta \pi_A( \vec  x)}\frac{\delta G}{\delta \phi^A(\vec  x)}\right)\,,
\ee
where we use the collective notation $\phi^A=(h_{ij}, N, N^i)$ and $\pi_A=(\pi^{ij}, \pi_N,\pi_i)$.

\subsection{Lagrangians up to $L_4$}
It is straightforward to apply the procedure outlined above to our Lagrangians up to $L_4$, because the expression (\ref{invert}) for $\pi^{ij}$ is linear in $K_{ij}$ and can be easily inverted. Including $L_5$ is more involved, as (\ref{invert}) is quadratic in $K_{ij}$ and we briefly discuss the procedure in the next subsection. 

Therefore, assuming that  $L_5$ is absent, i.e.~$A_5=B_5=0$, one can immediately invert (\ref{invert}) to find
\be
K_{ij}=-\frac{1}{A_4 \sqrt{h}}\left(\pi_{ij}-\frac12 \pi h_{ij}\right)-\frac{A_3}{4A_4}h_{ij}\,.
\ee
Using (\ref{extrinsic}), it is then straightforward to express $\dot h_{ij}$ as a function of $\pi_{ij}$ and to substitute the result in (\ref{Hamiltonian}). Using integrations by parts to get rid of the derivatives of the shift, one finds that the Hamiltonian can be written in the form
\be
H=\int d^3 \vec  x \left[ N {\cal H}_0(N)+N^i{\cal H}_i\right]\,,
\ee
with 
\begin{align}
{\cal H}_0\equiv &-\frac1{\sqrt{h} \Bfour } \bigg(\pi_{ij} \pi^{ij}-\frac12\pi^2 \bigg)-\frac{\Athree}{2\Bfour}\pi+\sqrt{h} \left(\frac{3\Athree^2}{8\Bfour} -\Atwo \right) - \sqrt{h}\, \Afour \R \;, \\
{\cal H}_i\equiv& -2  D_j \pi_{\ i}^j  \, .
\end{align}
As mentioned in the previous section, by specializing the above expressions to the case $B_4=-A_4=1/(16\pi G)$ and $A_2=A_3=0$ one recovers the usual general relativity Hamiltonian. In the general case, however, the $A_a$ and $B_a$ are functions  of $N$, so that ${\cal H}_0$ now depends on $N$, in contrast with general relativity. This difference plays a crucial role, as we will see below. 

Let us now consider the time evolution of  the primary constraints. Imposing that they are conserved in time  leads to the so-called  \emph{secondary constraints}. For the first constraint,  $\pi_N\approx 0$, one finds 
\be
\dot\pi_N = \big\{\pi_N ,\tH\big\} \approx \big\{\pi_N  ,H\big\}=-\frac{\partial }{\partial N } \left(N{\cal H}_0\right) \;,
\ee
where the symbol $\approx$ denotes equality in a ``weak'' sense, i.e.~restricted to the constrained phase space.
Thus, the above equation yields the secondary constraint,
\be
\label{tilde_H0}
\tilde{\cal H}_0\equiv{\cal H}_0+N\frac{\partial {\cal H}_0}{\partial N} \approx 0\,.
\ee
Note that, in general relativity, ${\cal H}_0$ is independent of $N$, thus leading to the familiar Hamiltonian constraint $\tilde{\cal H}_0={\cal H}_0\approx 0$.
Similarly, using 
\be
 \dot\pi_i = \big\{\pi_i,\tH\big\} \approx \big\{\pi_i  ,H \big\} =-{\cal H}_i\,,
\ee
the conservation in time of the three primary constraints $\pi_i\approx 0$ gives  the secondary constraints 
\be
{\cal H}_i \approx 0\,.
\ee 
These constraints are exactly the same as in pure general relativity, where they are associated with the invariance under spatial diffeomorphims.

Let us now compute the Poisson brackets of the constraints. We start with the constraints ${\cal H}_i $, for which the treatment is very similar to general relativity. It is convenient to introduce  the ``momentum'' function
\be
\M_f\equiv\int d^3 \vec x\,  f^i (\vec x) \, {\cal H}_i(\vec x)\,,
\ee
where the $f^i$ are three arbitrary functions of space. By reproducing the general relativity calculations (see {\em e.g.}~the appendix of \cite{Khoury:2011ay}), one finds
\be
\label{Poisson_M_M}
\left\{\M_f,\M_g\right\}=\M_h, \qquad h^i\equiv f^kD_k g^i-g^kD_k f^i\,.
\ee
It is also straightforward to check that 
\beq
\label{Poisson_M_T}
\left\{\M_f,\T_g\right\}=-\int d^3 \vec x \, g\, D_i(\T f^i)
=\int d^3 \vec x \, \T f^i D_i g\,,
\eeq
with 
\be
\T_g\equiv\int d^3 \vec  x\,  g(\vec x) \, \T(\vec x)\,,
\ee
where $g$ is an arbitrary function of space and $\T$ is any combination of the Hamiltonian that depends on $\pi^{ij}$ and $h_{ij}$, but \emph{not} on $N$. So $\T$ can be any of the following expressions,
\be
\T_1=\frac{1}{\sqrt{h}}\big(\pi_{ij} \pi^{ij}-\frac12\pi^2 \big)\,, \qquad \T_2=\pi\,, \qquad \T_3=\sqrt{h}\,,\qquad \T_4=\sqrt{h} R\,,
\ee
or any linear combination of these with coefficients \emph{independent} of $N$. In particular, (\ref{Poisson_M_T}) implies that in general relativity, where the constraint ${\cal H}_0$ does not depend on $N$, the Poisson bracket of $\M_f$ with ${\cal H}_0$ weakly vanishes.

If the combination $\T$ is now multiplied by a function of $N$, 
\beq
\tilde\T={\cal F}(N)\,  \T\,,
\eeq
one immediately deduces from (\ref{Poisson_M_T}) that 
\be
\big\{\M_f,\tilde \T_g\big\}=-\int d^3 \vec x \, g\, {\cal F}\,  D_i(\T f^i) \;,
\ee
and $\tilde \T$ cannot appear after integration by parts. However, by introducing the slightly modified constraints\footnote{Note that its form is similar to the total momentum constraint that would arise in general relativity with a scalar field.}
\be
{\tilde \H}_i \equiv  \H_i+\pi_N\partial_i N,
\ee
one obtains
\beq
\label{Poisson_M_Ttilde}
\big\{\tilde \M_f,\tilde \T_g\big\}=\big\{\M_f,\tilde \T_g\big\}-\int d^3 \vec x \, g\, \frac{\partial {\cal F}}{\partial N}
\T f^i D_iN=
-\int d^3 \vec x \, g\,  D_i(\tilde \T f^i)=\int d^3 \vec x \, \tilde\T f^i \, D_i g\,,
\eeq
where now $\tilde \T$ appears explicitly. 

This treatment also applies to any linear combination of $\tilde \T$ terms. In particular, it applies to ${\cal H}_0$, since this is given by a linear combination of ${\cal T}_a$  with coefficients that depend on time and $N$, and as a consequence it applies to $\tilde {\cal H}_0$ defined in eq.~\eqref{tilde_H0}.
Thus, from the above analysis one concludes that  the Poisson brackets of the constraints ${\tilde \H}_i$ with $\tilde \H_0$ vanish weakly, i.e. 
\be
\big\{ {\tilde \H}_i, {\tilde \H}_0\big\}\approx 0\,.
\ee
Using eq.~(\ref{Poisson_M_M}) and the fact that ${\H}_i$ does not depend on $N$, $\pi_N$, $N^i$ or $\pi_i$, it is also immediate to verify that 
\be
\label{Poisson_Hi}
\big\{ {\tilde \H}_i, {\tilde \H}_j\big\}\approx 0\,, \qquad \big\{ {\tilde \H}_i, \pi_N\big\}\approx 0, 
\qquad \big\{ {\tilde \H}_i, \pi_j\big\}\approx 0\,.
\ee
Therefore, the Poisson brackets of  the three constraints ${\tilde \H}_i$ with all the other constraints vanish weakly. The same is true for the three primary constraints $\pi_i \approx 0$.  Consequently, these  six constraints, associated with the 3-dimensional diffeomorphism invariance, are \emph{first-class} constraints. 

The remaining constraints, ${\tilde \H}_0$ and $\pi_N \approx 0$, satisfy the relations
\be
\label{tilde_H0_N}
\big\{ \pi_N(x), \pi_N(y)\big\}=0,\qquad \big\{\tilde{\cal H}_0,\pi_N\big\}=\frac{\partial\tilde{\cal H}_0}{\partial N}=2\frac{\partial{\cal H}_0}{\partial N}+\frac{\partial^2{\cal H}_0}{\partial N^2} \;.
\ee
 Provided that the derivative of $\tilde{\cal H}_0$ with respect to $N$ does not vanish, this shows that  these two constraints are  of the \emph{second-class} type,  in contrast with general relativity. 
 
 It is also useful to check that no additional constraint arises from the time evolution of the secondary constraints. Indeed, since
\be
\frac{d}{dt} \tilde{\cal H}_0= \frac{\partial \tilde{\cal H}_0}{\partial t}+\big\{\tilde{\cal H}_0, H\big\}+\lambda_N\frac{\partial\tilde{\cal H}_0}{\partial N}\,,
\ee
imposing the conservation of $\tilde{\cal H}_0$  simply fixes the Lagrange multiplier $\lambda_N$ without generating any new constraint, provided $\partial{\cal H}_0/{\partial N}$ does not vanish, which is assumed here. 
As for the momentum constraints, we simply have 
\be
\frac{d}{dt} \tilde{\cal H}_i=\big\{\tilde{\cal H}_i, H\big\}\approx 0\,,
 \ee
because the brackets of $\tilde{\cal H}_i$ with all the elements in $H$ vanish weakly, according to (\ref{Poisson_M_Ttilde}) and the first relation in (\ref{Poisson_Hi}).
 
In conclusion, we find that the dynamical system is, in general, characterized  by a 20-dimensional phase space with six first-class constraints and two second-class constraints. Each first-class constraint eliminates two canonical variables and each second-class constraint eliminates one canonical variable. In total, 14 canonical variables can be eliminated, which corresponds to a 6-dimensional physical  phase space, i.e.~three degrees of freedom. The difference with general relativity,  where all eight constraints are first-class thus leaving only two physical degrees of freedom, is due to the presence of a preferred slicing defined by the scalar field,  which breaks the full spacetime diffeomorphism invariance.

Let us briefly discuss  a special case where the second Poisson bracket in  (\ref{tilde_H0_N}) vanishes weakly,
which happens  when the whole $N$ dependence factorizes in  ${\cal H}_0$. Let us illustrate this case by considering  the Lagrangian $L_4$ with 
\be
B_4=-\frac{1}{A_4}\,.
\ee
In this case 
\be
{\cal H}_0=B_4\, \Big[\frac{1}{2 \sqrt{h} } \big(2\pi_{ij} \pi^{ij}-\pi^2 \big)-\sqrt{h} R\Big]\,
\ee
and 
\be
\tilde{\cal H}_0=\left(B_4+\frac{\partial B_4}{\partial N}\right)\, \Big[\frac{1}{2 \sqrt{h} } \big(2\pi_{ij} \pi^{ij}-\pi^2 \big)-\sqrt{h} R\Big]\,.
\ee
One then notices that the system is equivalent to general relativity, up to the redefinition of a new lapse function $\tilde N\equiv N B_4$. 

Finally, let us make a few considerations on the restriction to the unitary gauge which is at the basis of the Hamiltonian analysis of this section.  
An explicit Hamiltonian analysis without fixing unitary gauge seems to be a very tedious task in view of the complicated expressions of our theories in the covariant form, eqs.~\eqref{L22}-\eqref{L55}. Indeed, resorting to the unitary gauge has the huge advantage to hide the scalar degree of freedom in the metric and to enormously simplify the analysis. Thus, the full Hamiltonian treatment in an arbitrary gauge is beyond the scope of the present work. Fortunately, in Sec.~\ref{D} we present a completely different approach, which shows that the higher-order time derivatives in the equations of motion can be eliminated by using constraints that follow from these equations. This other approach is valid in any gauge and it confirms that no additional degree of freedom is necessary to describe higher-order time derivatives.
 
\subsection{Including the Lagrangian $L_5$}

The inclusion of $L_5$ makes the Hamiltonian analysis more involved, the main subtlety in this case being inverting eq.~\eqref{invert} in order to obtain $K_{ij}$ as a function of $\pi^{ij}$. However, this  technical difficulty  does not impair the basic counting of degrees of freedom, which is the main target of our Hamiltonian analysis. 

In the case when only $A_5$  is considered, from the last line of~\eqref{Lagrangian_uni} we obtain 
\begin{equation} \label{inv-L5}
\pi^{ij} \ = \  \frac{3 \sqrt{h} A_5}{2} \left[(K^2 - K_{mn} K^{mn}) h^{ij} + 2 (K^i_{\ l} K^{lj} - K K^{ij})\right]\, .
\end{equation}
Inverting the above equation is technically more involved and because $K_{ij}$ is essentially a ``square root" of $\pi_{ij}$ there is generally more than one branches of solutions. However, the inversion problem is well-defined locally around some non-singular chosen value of $K_{ij}$. It is worth mentioning how the problem can be tackled in practice with a systematic series expansion around,  for instance, a spatially flat Friedmann-Lema\^itre-Robertson-Walker (FLRW) configuration, 
\begin{equation}
(K_0)_i^{\ j} = H \delta_i^j, \qquad (\pi_0)_i^{\ j} = {3 \sqrt{h} A_5} H^2 \delta_i^j\, .
\end{equation}
We can then fix whatever value of the conjugate momentum through the new ``shifted" variable $\hat \pi_i^{\ j}$,
\begin{equation}
\pi_i^{\ j} \equiv (\pi_0)_i^{\ j} +  \frac{3 }{2 } \sqrt{h} A_5\, \hat \pi_i^{\ j},
\end{equation}
write a formal power expansion for $K_i^{\ j}$,
\begin{equation}
K_i^{\ j} \ = ( K_0)_i^{\ j} + \ku_i^{\ j} + \kd_i^{\ j} + \dots
\end{equation}
and solve~\eqref{inv-L5} order by order. By doing this, we obtain the recursive relations
\begin{align}
\ku_i^{\ j} \ &= \ - \frac{1}{2H} \left(\hat \pi_i^{\ j} - \frac{\hat\pi}{2}\delta_i^{j} \right),\\
\kd_i^{\ j} \ &= \ \frac{1}{4H}\left[\left(\ku^2 - \ku_m^{\ n} \ku_n^{\ m}\right) \delta_i^j + 4  \left( \ku_i^{\ l} \ku_l^{\ j} - \ku \ku_i^{\ j} \right) \right]\,,\dots \; ,
\end{align}
where $(K_a) \equiv (K_a)_i^{\ i}$.

A completely analogous procedure applies to other cases, such as when the full battery of terms is present, as in eq.~\eqref{invert}. In this case, the easily invertible part ($L_2$-$L_4$) can be used as the zeroth order piece and one can make a formal Taylor expansion in $A_5$.

\subsection{Generalizations} \label{sub_considerations}
Although we have focused our discussion on a specific class of theories, which represent a natural extension of Horndeski theories from the ADM point of view,  similar conclusions can be drawn for a much wider class of models.  Essentially, 
the basic ingredients that lead us to exclude the presence of unwanted additional degrees of freedom can be formulated in unitary gauge as 
\begin{enumerate}
\item unbroken spatial diffeomorphism (producing three first-class momentum constraints as in general relativity); 
\item absence of time derivatives of the lapse function $N$ (which makes the Hamiltonian constraint an \emph{algebraic} equation for $N$); 
\item absence of time derivatives of the extrinsic curvature $K_{ij}$ (which prevents that the Lagrangian depends on the ``accelerations", \emph{i.e.} the second time derivatives of $h_{ij}$).
\end{enumerate}
Such an approach  has already been used in the past to study, for instance, the behavior of specific models of  Horava's gravity \cite{Blas:2010hb}. In analogy with  Horava's gravity, one could consider  various combinations of the intrinsic curvature tensor and its spatial derivatives, as well as various combinations of the extrinsic curvature tensor,  as  recently discussed in \cite{Gao:2014soa}. Note, however, that these theories do not generically have the same decoupling limit as Horndeski, as it is the case for G${}^3$ theories (see discussion at the end of Sec.~\ref{G3}).

%%%%%%%%%%%%%%%%%%%%%%%%%%%%%%%%%%%%%%%%%%%%%%%%%%%%%%%%%%%%%%%%%%%%%%%%%
%%%%%%%%%%%%%%%%%%%%%%%%%%%%%%%%%%%%%%%%%%%%%%%%%%%%%%%%%%%%%%%%%%%%%%%%%
\section{Linear theory and coupling with matter}
\label{M}
%%%%%%%%%%%%%%%%%%%%%%%%%%%%%%%%%%%%%%%%%%%%%%%%%%%%%%%%%%%%%%%%%%%%%%%%%
%%%%%%%%%%%%%%%%%%%%%%%%%%%%%%%%%%%%%%%%%%%%%%%%%%%%%%%%%%%%%%%%%%%%%%%%%

The Hamiltonian analysis excludes the presence of extra degrees of freedom. However, one still needs to check that the remaining scalar and tensor degrees of freedom are not themselves  ghosts. In this section we compute the quadratic action for the perturbations of the propagating degrees of freedom and derive the conditions for which the kinetic terms have the right signs. We then add matter fields minimally coupled to gravity and study the phenomenology on small scales. We first perform this analysis in unitary gauge and then in Newtonian gauge. 

\subsection{Unitary gauge}
\label{M1}
Let us expand  action \eqref{full_action} around a spatially flat FLRW metric  following the general procedure developed in \cite{GLPV,PV} (see also \cite{Maldacena:2002vr}). We use the $\zeta$-gauge and write the spatial metric as 
\be
h_{ij} = a^2(t)e^{2 \zeta} (\delta_{ij} + \gamma_{ij} ) \;, \qquad \gamma_{ii} = 0 = \partial_i \gamma_{ij}\; ,
\ee
and we split the shift as 
\be
N^i = \partial_i \psi + N_V^i \;, \qquad \partial_i N_V^i=0 \;.
\ee 
Moreover, it is convenient to express the dependence of the second-order action on the function $A_a$ and $B_a$ introduced in the Lagrangians \eqref{Lagrangian_uni} in terms of  the following functions evaluated on the background,\footnote{The first four functions in eq.~\eqref{ABparameters} have been introduced by Bellini and Sawicki in Ref.~\cite{Sawicki}, where they consider linear perturbations in Horndeski theories, with the difference $\alpha_B^{\rm here}=-\alpha_B^{\rm there}/2$, which simplifies further the equations. In particular,
 $M^2$, $\alpha_K$, $\alpha_B$ and $\alpha_T$ respectively parameterize the effective Planck mass, a modification of the scalar kinetic term \cite{ArmendarizPicon:2000dh,ArmendarizPicon:2000ah}, a kinetic mixing  between the scalar and the metric (the so-called braiding)  \cite{Creminelli:2006xe,Creminelli:2008wc,Deffayet:2010qz,Pujolas:2011he} and a tensor speed excess. As stressed in such a reference and also shown in Appendix~\ref{app:BB}, these functions are just a convenient basis of the parameters previously introduced in the context of the so-called Effective Field Theory of Dark Energy in Refs.~\cite{EFTOr,Bloomfield:2012ff,GLPV,Bloomfield:2013efa} (see \cite{PV,Tsujikawa:2014mba} for reviews). Here we adopt this parameterization because it simplifies the notation. We also introduce a new function, $\alpha_H$, which parametrizes the deviation from Horndeski theories \cite{GLPV,Gleyzes:2014dya}.}
\be
\label{ABparameters}
\begin{split}
M^2 & \equiv -2 (A_4 + 3 H A_5) \, , \\
\alpha_K & \equiv - \frac{2 A_2 ' + A_2'' + 3 H (2 A_3' +A_3'')  + 6 H^2 (2 A_4' + A_4'') + 6 H^3( 2 A_5' + A_5'')}{2 H^2 (A_4 + 3 H A_5) }\, , \\  
\alpha_B & \equiv  -\frac{  A_3' + 4 H A_4' + 6 H^2 A_5'}{ 4H (A_4 + 3 H A_5)  }\, , \\
\alpha_T  & \equiv  - \frac{B_4 + \dot B_5/2}{A_4 + 3 H A_5} - 1 \;, \\
\alpha_H & \equiv  - \frac{B_4 + B_4'  - H B_5'/2 }{A_4 + 3 H A_5} -1 \;,
\end{split}
\ee
where a prime denotes a derivative with respect to $N$ and a dot a derivative with respect to $t$. We discuss in Appendix~\ref{app:BB} how these functions are related to the general formalism of Ref.~\cite{GLPV}.

Higher (spatial) derivative terms proportional to $(\partial^2 \psi)^2$, which are contained in quadratic products of the extrinsic curvature, cancel from the action up to a total derivative because of the particular combinations in which these products appear in eq.~\eqref{Lagrangian_uni}. By varying the quadratic action with respect to $N^i$, one obtains the momentum constraints, whose solution is $N_V^i =0$ and
\be
 N = 1+  \frac{1}{1+ \alpha_B } \frac{ \dot \zeta}{H} \;. \label{mc}
\ee
After substitution of this equation  
 into the quadratic action, 
all the terms containing $\psi$  drop out, up to total derivatives \cite{PV}. For this reason, we do not need the Hamiltonian constraint, obtained by varying the action with respect to $N$, to solve for $\psi$.  After some manipulations the quadratic action becomes \cite{GLPV,PV,Sawicki}
\be 
\label{lag-quad}
S^{(2)}= \frac12 \int d^4 x \, a^3  \bigg[ {\cal L}_{\dot \zeta \dot \zeta} \dot \zeta^2  +  {\cal L}_{\partial \zeta \partial \zeta} \frac{(\partial_i \zeta)^2}{a^2}    + \frac{M^2}{4} {\dot \gamma}_{ij}^2 - \frac{M^2}{4}  (1+ \alpha_T) \frac{(\partial_k \gamma_{ij})^2}{a^2}   \bigg]\;,
\ee
where 
\begin{align}
{\cal L}_{\dot \zeta \dot \zeta} &\equiv   M^2  \frac{ \alpha_K +6 \alpha^2_B }{(1 + \alpha_B)^2}  \;, \\
{\cal L}_{\partial \zeta \partial \zeta} &\equiv 2  M^2 (1+\alpha_T) - \frac{2}{a} \frac{d}{dt} \bigg[ \frac{a M^2 (1+ \alpha_H) }{H(1+ \alpha_B)} \bigg]  \;.\end{align}
As expected from the previous Hamiltonian analysis, the quadratic Lagrangian~\eqref{lag-quad} does not contain higher-order time derivatives. 
As a consequence of the particular combination of extrinsic curvature in eq.~\eqref{Lagrangian_uni}, neither does it contain higher space derivatives. 

The condition required to  ensure that the  propagating degrees of freedom are not ghost-like is that their time kinetic terms are positive, ${\cal L}_{\dot \zeta \dot \zeta} > 0$ and $M^2 > 0$.
Moreover, gradient instabilities are avoided when the speed of sound of the scalar and tensor propagating degrees of freedom,
\be
\label{css}
c_s^2 \equiv - \frac{{\cal L}_{\partial \zeta \partial \zeta}}{{\cal L}_{\dot \zeta \dot \zeta}} \;, \qquad c_{_T}^2 \equiv 1+\alpha_T \;,
\ee
are also positive, $c_s^2 >0 $ and $c_{_T}^2 >0 $.

\subsection{Adding matter: $P(\sigma, Y)$}
\label{M2}

To study our theories in the presence of matter fields minimally coupled to gravity, we add to action \eqref{full_action} 
a $k$-essence type action describing a matter scalar field  $\sigma$ (not to be confused with the  dark energy field $\phi$),
\be
S_m=\int d^4x\sqrt{-g}\;P(\Y,\sigma)\,, \qquad  \Y \equiv g^{\mu \nu}\partial_\mu \sigma \partial_\nu \sigma\,,
\ee
with sound speed $c_m^2 \equiv {P_\Y}/({P_\Y-2\dot{\sigma}_0^2 P_{\Y \Y}})$. 

We can then expand at second order these actions and repeat the procedure discussed earlier. 
To describe matter fluctuations it is convenient to use  
the gauge-invariant variable
 $Q_\sigma \equiv \delta \sigma  - ({\dot \sigma_0}/{H})\zeta $.
After substitution of the momentum constraints, the final action expressed in terms of $\zeta$ and $Q_\sigma$ reads
\begin{align}
 {S}^{(2)} = \int d^4 x a^3 & \bigg[  \frac12 \bigg(\tilde {\cal L}_{\dot \zeta \dot \zeta}  \dot \zeta^2  +  \tilde {\cal L}_{\partial \zeta \partial \zeta}\frac{(\partial_i \zeta)^2}{a^2} \bigg)  - \frac{P_\Y}{c_m^2}  \bigg( \dot Q_\s^2-c_m^2 \frac{(\partial_i Q_\s)^2}{a^2} \bigg) \nonumber \\
&
 - \frac{2 \dot{\sigma}_0P_\Y}{ H c_m^2 (1+ \alpha_B)} \bigg( \alpha_B \dot\zeta \dot Q_\s - c_m^2 ( \alpha_B-\alpha_H ) \frac{\partial_i \zeta\partial_i Q_\s}{a^2} \bigg)\nonumber \\
&
  + m_\zeta^2 \zeta^2 + m_\sigma^2 Q^2_\sigma + m_c^2 \zeta Q_\sigma + \lambda \dot \zeta Q_\sigma \bigg]
\;,
\end{align}
with the new coefficients for the kinetic and gradient terms of $\zeta$
\begin{align}
\tilde {\cal L}_{\dot \zeta \dot \zeta} & =   {\cal L}_{\dot \zeta \dot \zeta} +  \frac{ \rho_m + p_m }{  H^2 c_m^2  } \, \bigg( \frac{\alpha_B}{1+\alpha_B} \bigg)^2  \;, \\
\tilde {\cal L}_{\partial \zeta \partial \zeta} & = {\cal L}_{\partial \zeta \partial \zeta}  -  \frac{ \rho_m + p_m }{  H^2  } \, \bigg(1 -  \frac{ 2 (1+ \alpha_H)}{1 + \alpha_B} \bigg) \;,
\end{align}
where we have used $2 \dot \sigma_0^2P_\Y = -(\rho_m +p_m)$. The second line contains two derivative couplings between $\zeta$ and $Q_\sigma$ while the third line contains non-derivative terms, which are irrelevant for the present discussion.

The kinetic matrix for $(\zeta, Q_\s)$ reads
\be \label{matkine}
\mathcal{M} = \,
\frac12 \begin{pmatrix}
\tilde {\cal L}_{\dot \zeta \dot \zeta} \omega^2 +\tilde {\cal L}_{\partial \zeta \partial \zeta}  k^2 & A \big[\alpha_B \omega^2 - c_m^2 (\alpha_B - \alpha_H) k^2\big] \\
A  \big[\alpha_B \omega^2  - c_m^2 (\alpha_B - \alpha_H) k^2\big] & - 2 P_Y c_m^{-2} ( \omega^2 - c_m^{2} k^2) 
\end{pmatrix} \, , \qquad A = -\frac{2 \dot{\sigma}_0P_\Y}{ H c_m^2 (1 + \alpha_B)} \;.
\ee
Requiring that its determinant  vanishes yields the dispersion relation
\begin{equation}
(\omega^2 - c_m^2 k^2) ( \omega^2 - \tilde c_s^2 k^2)  =  (c_s^2 - \tilde c_s^2) \, \bigg( \frac{\alpha_H}{ 1+\alpha_H  } \bigg)^2  \, \omega^2 k^2 \, , \label{km} 
\end{equation}
with
\be
\label{tildecss}
\tilde c_s^2 \equiv  c_s^2 -  \frac{  \rho_m +p_m}{ H^2 M^2} \,  \frac{(1+\alpha_H)^2}{  \alpha_K +6 \alpha_B^2  } \;.
\end{equation}
From this equation one derives the two dispersion relations $\omega^2 = c^2_{\pm} k^2$.  For Horndeski theories ($\alpha_H =0$), the matter sound speed is unchanged, despite the  presence of couplings in the action between the time and space derivative of $\zeta$ and $Q_\sigma$, i.e.~the non- vanishing of the non-diagonal terms in the kinetic matrix. Indeed, these  couplings are precisely proportional to $ \omega^2 - c_m^2 k^2$ and give the standard dispersion relation for matter. However, this is no longer true with our non-Horndeski extensions,  where $\alpha_H \neq 0$.

%%%%%%%%%%%%%%%%%%%%%%%%%%%%%%%%%%%%%%%%%%%%%%%%%%%%%%%%%%%%%%%%%%%%%%%%%
%%%%%%%%%%%%%%%%%%%%%%%%%%%%%%%%%%%%%%%%%%%%%%%%%%%%%%%%%%%%%%%%%%%%%%%%%
\subsection{Newtonian gauge}
%%%%%%%%%%%%%%%%%%%%%%%%%%%%%%%%%%%%%%%%%%%%%%%%%%%%%%%%%%%%%%%%%%%%%%%%%
%%%%%%%%%%%%%%%%%%%%%%%%%%%%%%%%%%%%%%%%%%%%%%%%%%%%%%%%%%%%%%%%%%%%%%%%%
\label{HEsect}

We now study linear perturbations for our theories in the presence of a more general type of matter by considering a gauge often employed in the study of  cosmological perturbations: the Newtonian gauge, where   the metric reads
\be
\label{NG}
ds^2 = - (1+2 \Phi) dt^2+ a^2 (1-2\Psi) d \vec x^2 \;,
\ee
taking into account only scalar perturbations. 

Let us directly expand the action for the sum of the Lagrangians \eqref{L22}--\eqref{L55}  up to quadratic order around the background field solution $\phi_0(t) = t$, i.e.,
\be
\phi = t + \pi (t, \vec x) \;,
\ee
where $\pi$ describes the scalar field perturbation.\footnote{Assuming a monotonic $\phi_0 = \phi_0(t)$, one can always make a field redefinition of $\phi$ and choose the background solution $\phi_0 = t$.} The quadratic action for linear perturbations is  given by
\be
\begin{split}
\label{KinetAct}
S=& \!\int \!d^4x a^3M^2 \bigg\{ \frac12 H^2 \alpha_K \pid^2+\bigg[\dot H  + \frac1{2M^2} \big(\rho_m + p_m +2(M^2 H \alpha_B)^{\hbox{$\cdot$}} -2(H M^2 \alpha_H)^{\hbox{$\cdot$}} \big)+  H^2 (\alpha_B - \alpha_M) \\ &+ H^2 (\alpha_T - \alpha_H)      \bigg] \frac{(\gpi)^2}{a^2}
-3 \dot \Psi^2 + (1+\alpha_T) \frac{(\nabla\Psi)^2}{a^2} 
+2 H ( \alpha_B - \alpha_H )\nabla\Phi\gpi \\ 
& -2 H (\alpha_M -\alpha_T) \frac{ \nabla\Psi \gpi}{a^2} 
+6 H  \alpha_B\pid \dot \Psi +H^2 ( 6 \alpha_B-\alpha_K ) \Phi \pid  -2(1+\alpha_H) \frac{\nabla\Phi \nabla\Psi}{a^2} \\
& - 6 H(1+ \alpha_B) \dot \Psi \Phi+ H^2 \left( \frac12 \alpha_K - 3 (1 +2 \alpha_B)  \right)\Phi^2
+2 \alpha_H \frac{\nabla\pid \nabla \Psi}{a^2} + \ldots \bigg\} \; , 
\end{split}
\ee 
where we have used the background equations to rewrite the coefficient of $(\nabla \pi)^2$.  We have written explicitly all the terms that are quadratic in derivatives, as well as  other terms involving 
$\Phi$ {\em without} derivatives because they also contribute to the kinetic limit as we will see below. The ellipses in the last line stand for  all the other terms, irrelevant for the present discussion.
As expected from the Lagrangians \eqref{L44} and \eqref{L55},  the quadratic action in the Newtonian gauge contains a higher order derivative term, $\nabla \dot \pi \nabla \Psi$, which is proportional to 
the non-Horndeski coefficient $\alpha_H$. 
This term  generates higher order (one time- and two spatial-) derivative terms  in the equations of motion, as discussed in detail in Ref.~\cite{GLPV}.

It is possible to find a  redefinition of the metric perturbations  that  de-mixes  the new metric variables  from the scalar field $\pi$ and removes the higher derivative term from the gravitational action. In Brans-Dicke theories such de-mixed variables are usually referred to as \emph{Einstein-frame} quantities. In our much more general framework they are explicitly given by
\be
\label{toE}
\begin{split}
\Phi_E & \equiv     \frac{1+\alpha_H}{1+\alpha_T}\Phi +   \bigg( \frac{1+\alpha_M}{1+\alpha_T} - \frac{1+  \alpha_B}{1+\alpha_H} \bigg) H \pi - \frac{\alpha_H}{{1+\alpha_T}} \pid  \; , \\
\Psi_E & \equiv \Psi+  \frac{\alpha_H-\alpha_B}{1+\alpha_H} H \pi \, .
\end{split}
\ee
Using this change of variables into the quadratic action, one ends up with 
\be
\begin{split}
\label{PiKinet1}
S =\int & d^4x a^3 M^2 \bigg\{ \frac{H^2 {\cal L}_{\dot \zeta \dot \zeta} }{2M^2} \bigg(\frac{1+ \alpha_B}{1+\alpha_H } \bigg)^2  \left(\pid^2-\tilde c_{s}^2\frac{(\gpi)^2}{a^2}\right) \\
&-3 \dot \Psi_E^2 +\frac{1+\alpha_T}{a^2} \big[ (\nabla\Psi_E)^2 -2\nabla\Phi_E\nabla\Psi_E \big]+\cdots \bigg\}\,,
\end{split}
\ee
whose first line corresponds to the action of a  minimally coupled scalar field.
In particular, the term proportional to $\dot \pi$ in the definition of $\Phi_E$  entails the  removal of  the higher derivative term $\nabla \dot \pi \nabla \Psi$.

Let us now consider matter. Since  it is minimally coupled to the original  metric, i.e.
\be
\label{actmat}
L_{\rm int} \equiv  \frac12 \delta g_{\mu\nu}\delta T^{\mu\nu} = - (\Phi \delta\rho_m+3\Psi\delta p_m)\;, 
\ee
it becomes coupled to $\pi$ after the field redefinition \eqref{toE}. When $\alpha_H = 0$, matter is  coupled to  the gravitational sector with standard terms, $\Phi_E \delta \rho_m$ and $\Psi_E \delta p_m$, as well as to $\pi$ via  fifth-force terms, $\pi \delta \rho_m$ and $ \pi \delta p_m$. These couplings can be neglected on scales smaller than the matter sound horizon, \emph{i.e.} for $k \gg Ha/c_m$, where $c_m$ is the matter sound speed. However, in the non-Horndeski case ($\alpha_H\neq 0$), the interaction Lagrangian \eqref{actmat} contains a new coupling proportional to the time derivative of the scalar $\pi$,
\be
L_{\rm int} \supset  - \frac{\alpha_H}{{1+\alpha_H}} \dot \pi \delta \rho_m\;,
\ee 
which cannot be neglected on scales smaller than the sound horizon. Indeed, 
on these scales,  the propagation equations for the  density contrast $\delta \rho_m$ and field perturbation $\pi$ become
\begin{align}
\ddot \delta \rho_m -c_m^2\frac{\nabla^2 \delta \rho_m}{a^2}  -(\rho_m+p_m) \frac{\alpha_H}{1+\alpha_H} \frac{\nabla^2 \pid}{a^2} & \approx 0 \; , \\
\ddot \pi-\tilde c_{s}^2\frac{\nabla^2 \pi}{a^2} -  \frac{1 }{H^2 {\cal L}_{\dot \zeta \dot \zeta}} \frac{\alpha_H(1+\alpha_H) }{(1+\alpha_B)^2} \dot{\delta\rho}_m & \approx 0\; , \label{sfcoup}
\end{align}
where the symbol $\approx$ stands for an equality in the kinetic limit. One can check that the propagation equation is given also in this case by eq.~\eqref{km}.
In contrast to the standard Jeans lore, the gravitational scalar mode $\pi$ cannot be decoupled from matter by going at sufficiently short distances. The presence of the scalar field perturbations impacts the propagation of matter fluctuations, by changing their sound speed.

 %%%%%%%%%%%%%%%%%%%%%%%%%%%%%%%%%%%%%%%%%%%%%%%%%%%%%%%%%%%%%%%%%%%%%%%%%
%%%%%%%%%%%%%%%%%%%%%%%%%%%%%%%%%%%%%%%%%%%%%%%%%%%%%%%%%%%%%%%%%%%%%%%%%
\section{Field redefinitions}
\label{D}
%%%%%%%%%%%%%%%%%%%%%%%%%%%%%%%%%%%%%%%%%%%%%%%%%%%%%%%%%%%%%%%%%%%%%%%%%
%%%%%%%%%%%%%%%%%%%%%%%%%%%%%%%%%%%%%%%%%%%%%%%%%%%%%%%%%%%%%%%%%%%%%%%%%

This section is devoted to exploring some mathematical properties of the class of theories that we are proposing and to confirm their soundness for subclasses of these theories. The approach discussed in this section does not rely on the ADM formulation and we do not need to assume $\nabla_\mu \phi$ being timelike, in contrast with our Hamiltonian analysis.

First, we analyse disformal transformations and focus on a specific class of disformal transformations that act as a ``morphism" on our theories, in the same way in which conformal transformations 
preserve the basic structure of Brans-Dicke theories. Next, we show how to relate, by means of such disformal transformations, subsets of our theories---i.e.~$L_4$ and $L_5$, separately studied in Secs.~\ref{sub_L4} and~\ref{sub_L5}, respectively---into Horndeski ones. 
As these disformal transformations conserve the number of degrees of freedom, this is yet another proof that  our theories do not contain ghosts, even if they contain higher derivatives.  In the cases in which the mapping with Horndeski is possible, we further clarify this issue 
in Sec.~\ref{sec-6:5}, by showing that naively higher-derivative equations can be reduced to second-order ones. In passing, we also verify in Sec.~\ref{sub_coupling} that the presence of matter does not spoil the soundness of the theory.

\subsection{Disformal transformations}
In this section we compute the transformation properties of our theories under \emph{disformal transformations}. 
More precisely, we consider a field redefinition of the metric tensor made of a conformal transformation and of a further lightcone structure-changing piece  \cite{Bekenstein:1992pj}, 
\begin{equation} \label{disformal}
g_{\mu \nu} \to \bar g_{\mu \nu} = \Omega^2(\phi,X)\,  g_{\mu \nu} + \ga(\phi,X)\,  \partial_\mu  \phi \, \partial_\nu \phi \, .
\end{equation}
For convenience, we directly work in  unitary gauge even though the same results can be reached using a covariant approach (see {\em e.g.}~\cite{DisInv,Zumalacarregui:2013pma}). As we shall see, the use of the unitary gauge considerably simplifies the calculations.

In this gauge, the dependence of $\Omega$ and $\ga$ on $\phi$ and $X$ translates into an explicit dependence on the time variable $t$ and on the lapse function $N$. Moreover, we choose time to coincide with $\phi$, so that $\partial_\mu \phi = \delta_\mu^0$
and eq.~\eqref{disformal} reads, in ADM components,
\begin{equation} 
\label{3-d}
\bar N^i  = N^i \;, \qquad 
\bar h_{ij} = \Omega^2 (t,N) \, h_{ij} \;, \qquad 
\bar N^2= \Omega^2(t,N) \, N^2 - \ga \, (t,N)\, . 
\end{equation}
Thus,  the volume element is transformed accordingly,
\be
\sqrt{- \bar g} = \sqrt{ -g} \, \Omega^3 \sqrt{\Omega^2 - \ga /N^2} \;.
\ee

In order to find how the three-dimensional Ricci scalars, $\R$ and $\Rb$,  are related to each other, we can apply the standard formulae to the conformal transformations of the 3-d metric~\eqref{3-d} (see {\em e.g.}~\cite{Wald:1984rg}),
\begin{equation} \label{mammate}
\Rb = \Omega^{-2}\left[\R - 4 D^2 \ln \Omega - 2 \partial_i (\ln \Omega) \partial^i (\ln \Omega)\right]\, .
\end{equation} 
Moreover, using the definition of the extrinsic curvature, eq.~\eqref{extrinsic}, one finds
\begin{equation}
\bar K_{ \ i}^j \ =\ \frac{N}{\bar N} \left[K^j_{\ i} \, - \,N  g^{0 \mu} \partial_\mu \ln \Omega \, \delta_{ \ i}^j \right]. 
\end{equation}

As in unitary gauge $\Omega$  depends on the spatial coordinates only through $N$, it 
makes a lot of difference whether or not $\Omega$ depends on $N$. If it does, the transformation \eqref{3-d} generates derivatives of $N$ explicitly  in the action, therefore changing the structure of action \eqref{full_action}. Thus, transformations with $\Omega$ dependent on $N$ do not preserve the G$^3$ form of the Lagrangian. 
On the contrary, if $\Omega$ is independent of $N$,
eq.~\eqref{3-d} is just an overall (spatial) coordinate-independent rescaling  from the 3-dimensional point of view and the structure of our theory does not change after the field redefinition.

Thus, let us consider an $N$ independent conformal factor, $\Omega = \Omega(t)$. 
Explicitly, starting from  the action \eqref{full_action} written in terms of the barred metric quantities with coefficients $\bar{A}_a$ and $\bar{B}_a$,  and making the substitution (\ref{3-d}) with $\Omega = \Omega(t)$, one ends up with an  
action in terms of the unbarred quantities. Remarkably, this new action shares the same structure \eqref{full_action}, up to a reshuffling of the coefficients:
\be
\begin{split}
\label{transf_disformal}
A_2 &= \frac{\Omega^{3} \bar N}{N} \left[ \bar A_2 +  3 \frac{d \ln  \Omega}{d \bar t} \bar A_3 +   6  \bigg( \frac{d \ln  \Omega}{d \bar t} \bigg)^2 \bar A_4  + 6 \bigg( \frac{d \ln  \Omega}{d \bar t} \bigg)^3 \bar A_5 \right]\, , \\
A_3 & = \Omega^{3} \left[\bar A_3 + 4 \frac{d \ln  \Omega}{d \bar t} \bar A_4 + 6 \bigg( \frac{d \ln  \Omega}{d \bar t} \bigg)^2 \bar A_5\right] \;,  \\
A_4 & =  \frac{\Omega^{3} N}{\bar N}  \left[ \bar A_4 + 3 \frac{d \ln  \Omega}{d \bar t}  \bar A_5 \right] \;, \\ 
A_5  & =  \frac{\Omega^3 N^2}{\bar N^2}  \bar A_5 \;, \\ 
B_4 & = \frac{\Omega \bar N}{N} \left[ \bar B_4 - \frac12 \frac{d \ln  \Omega}{d \bar t} \bar B_5\right] \;, \\
B_5 & = \Omega  \bar B_5\, ,
\end{split}
\ee
where $d \bar t \equiv \bar N dt$.
One notes that, in this disformal transformation,  a Lagrangian of a given order generally contributes also to the lower-order Lagrangians. For instance, the transformation of ${L}_4$ contains also $ {L}_3$ and $ {L}_2$ pieces. Only when $\Omega = $ const.~does this mixing not occur.

Although we have worked specifically in the unitary gauge, it is straightforward to perform the same analysis  covariantly, directly  with the 4-dimensional transformation 
\begin{equation} \label{disformal_G3}
g_{\mu \nu} \to \bar g_{\mu \nu} = \Omega^2(\phi)\,  g_{\mu \nu} + \ga(\phi,X)\,  \partial_\mu  \phi \, \partial_\nu \phi \,.
\end{equation}
One then obtains relations between the coefficients $A_a(\phi,X)$, $B_a(\phi,X)$ and $\bar A_a(\phi, \bar X)$, $\bar B_a(\phi, \bar X )$, which are essentially the above relations (\ref{transf_disformal}) with the correspondence
$N=1/\sqrt{-X}$ and $\bar N=1/\sqrt{-\bar X}$. The relation between $X$ and $\bar X$ can be computed by contracting the inverse metric,
\beq
\bar g^{\mu\nu}=\Omega^{-2}\left(g^{\mu\nu}-\frac{\ga}{\ga X+\Omega^2}\partial^\mu\phi\partial^\nu\phi\right)\;,
\ee
with  $\partial_\mu \phi \partial_\nu \phi$. This gives 
\beq
\bar X=\frac{X}{\ga X+\Omega^2}\,, \qquad X=\frac{\Omega^2 \bar X}{1- \ga \bar X}\,.
\eeq
We also have
\be
\label{sr}
\frac{\sqrt{-g}}{\sqrt{- \bar g}}=\frac{\sqrt{1-\bar X\ga}}{\Omega^4}=\frac{1}{\Omega^3 \sqrt{\ga X+\Omega^2}}\,,
\ee
which implies in particular that, in unitary gauge,
\be
\frac{N}{\bar N}=\frac{\sqrt{1-\bar X\ga}}{\Omega}=\frac{1}{\sqrt{\ga X+\Omega^2}}\,,
\ee
which can be substituted in eq.~\eqref{transf_disformal}.

\subsection{Link between $L_4$ and   Horndeski} \label{sub_L4}
The disformal transformations discussed in the previous subsection can be used to relate Horndeski theories with our general Lagrangians. 

First, let us start from a Horndeski Lagrangian $L_4^H$ expressed in terms of the metric $\bar{g}_{\mu\nu}$, with coefficients $\bar{A}_4( \phi, \bar X)$ and $\bar{B}_4( \phi, \bar X)$ satisfying the Horndeski condition (see eq.~\eqref{gal})
\be
\label{horndeski_L4}
\bar A_4=-\bar B_4+2\bar X \bar B_{4\bar X}\,. 
\ee
Substituting in this Lagrangian  the expression
\be
\label{disformal_L4}
\bar g_{\mu \nu}  =  g_{\mu \nu} + \ga_4( \phi,  X)\,  \partial_\mu   \phi \, \partial_\nu  \phi \;,
\ee
 leads to a G$^3$ Lagrangian, now expressed in terms of the metric $g_{\mu\nu}$ and $X$, with coefficients
$A_4(\phi, X)$ and $B_4(\phi,X)$. According to the results of the previous subsection, specialized to the case $\Omega=1$, the link between the old and new coefficients is given by the relations
\be
\label{barA4toA4}
\bar A_4 ( \phi, \bar X) ={A_4 (\phi, X)} {\sqrt{1+  X\ga_4 }}\,, \qquad  A_4 ( \phi,  X) ={\bar A_4 (\phi, \bar X)} {\sqrt{1- \bar  X\ga_4 }} 
\ee
and 
\be
\label{barB4toB4}
\bar B_4 ( \phi, \bar X)= \frac{B_4 (\phi, X)}{\sqrt{1+  X\ga_4 }} \,, \qquad
B_4 ( \phi, X)= \frac{\bar B_4 (\phi, \bar X)}{\sqrt{1-\bar   X\ga_4 }}\;,
\ee
with
\beq
\bar X=\frac{X}{1+\ga_4 X}\,, \qquad X=\frac{\bar X}{1-\ga_4 \bar X}\,.
\eeq
The Horndeski condition (\ref{horndeski_L4}) on the coefficients $\bar A_4$ and $\bar B_4$ implies the following relation between $\ga_4$ and the new coefficients $A_4$ and $B_4$:
\beq
\label{Lambda_4X}
\ga_{4X}=\frac{A_4+B_4- 2X B_{4X}}{X^2 A_4}\,.
\eeq
It is thus clear that the new Lagrangian, expressed in terms of the metric $g_{\mu\nu}$, is not of the Horndeski type unless $\ga_4$ is independent of  $X$. This is consistent with the findings of Ref.~\cite{DisInv} that  the  Horndeski form of the Lagrangian is preserved under a restricted version of \eqref{disformal_G3}, in which the disformal function $\ga$, like $\Omega$, does not depend on $X$.

Conversely, if we start with a G$^3$ Lagrangian without $L_5$ terms, but otherwise with arbitrary functions 
$A_4(\phi, X)$ and $B_4(\phi,X)$, one can always rewrite it as a Horndeski Lagrangian $L_4^H$, provided that the transformation function $\ga_4$ is a solution of the differential equation (\ref{Lambda_4X}). 
Note that the  field redefinition \eqref{disformal_L4} is {\em well-defined}, in the sense that it leaves invariant the number of degrees of freedom (see other examples in \cite{deRham:2011qq}). Indeed, one can express $g_{\mu \nu}$ in terms of $\bar g_{\mu\nu}$ and $\phi$ without introducing additional degrees of freedom. As the set of fields $(\bar g_{\mu \nu}, \phi)$ obeys the Horndeski equations of motion, it describes three degrees of freedom. By the field transformation \eqref{disformal_L4}, also $( g_{\mu \nu}, \phi)$ obeying the equations of motion derived from the G$^3$ Lagrangian $L_4$ describe the same number of degrees of freedom, i.e.~three.
This essentially confirms the Hamiltonian analysis of Sec.~\ref{H} which excludes the presence of more than three degrees of freedom in G$^3$ theories. 
As expected, the field redefinition \eqref{disformal_L4} partly de-mixes the metric and scalar field kinetic mixing presented in Sec.~\eqref{M}. In Newtonian gauge, this corresponds to removing the higher-derivative coupling $2 \alpha_H {\nabla\pid \nabla \Psi}$ from action \eqref{KinetAct}, as explicitly shown in Appendix~\ref{app:CC}.

\subsection{Link between $L_5$ and   Horndeski} \label{sub_L5}

The same procedure described above applies to $L_5$ Lagrangians along similar lines. Namely, one can always relate a G$^3$ Lagrangian with arbitrary $A_5$ and $B_5$, but with $A_4=B_4=0$, to a Horndeski Lagrangian of the type $L_5^H$, provided the two metrics are related by 
\be
\bar g_{\mu \nu}  =  g_{\mu \nu} + \ga_5( \phi,  X)\,  \partial_\mu   \phi \, \partial_\nu  \phi \;,
\ee
with $\ga_5$ satisfying the condition 
\beq
\label{Lambda_5X}
\ga_{5X}=\frac{3A_{5}+XB_{5X}}{3 X^2 A_5} \;.
\eeq
Analogously to the above discussion, this follows from requiring that $\bar A_5$ and $\bar B_5$, given by (see eq.~\eqref{transf_disformal})
\be
\bar A_5 ( \phi, \bar X) ={A_5 (\phi, X)} {(1+  X\ga_5 )}\,, \qquad  \bar B_5 ( \phi, \bar X)= {B_5 (\phi, X)} \,, 
\ee
satisfy  Horndeski condition (see eq.~\eqref{gal}),
\be
\label{horndeski_L5}
 \bar A_5 = - \bar X \bar B_{5 \bar X}/3\;. 
\ee

However, one cannot in general re-express an arbitrary G$^3$ Lagrangian as a Horndeski Lagrangian via a disformal transformation, because the would-be transformation coefficient $\ga$ cannot satisfy simultaneously the two differential equations (\ref{Lambda_4X}) and (\ref{Lambda_5X}).

\subsection{Coupling to matter} \label{sub_coupling}
When the G$^3$ Lagrangian can be re-expressed as a Horndeski Lagrangian, i.e.~in either of the two cases discussed above, the coupling between  matter and the gravitational sector, now described by $\bar g_{\mu\nu}$ and $\phi$,  becomes more complicated since the matter Lagrangian depends on the combination
\be
g_{\mu\nu}=\bar g_{\mu\nu}-\ga (\phi, \bar X) \partial_\mu \phi\partial_\nu\phi,
\ee
or its inverse,
\be
g^{\mu\nu}=\bar g^{\mu\nu} +\frac{\Gamma(\phi,\bar X)}{1-\Gamma(\phi,\bar X)\bar X}\bar g^{\rho \mu} \bar g^{\sigma \nu} \partial_\rho \phi\partial_\sigma \phi\,.
\ee
 Let us illustrate this with the simple example of an ordinary matter scalar field, minimally coupled to the metric $g_{\mu\nu}$. Its action, which intially reads 
 \be
S_{\text{mat}}=\int d^4x  \sqrt{-g} \left[-\frac12{ g^{\mu\nu}\partial_\mu \sigma \partial_\nu \sigma} -V(\sigma)\right]\, ,
\ee
becomes, when expressed in terms of $\bar g_{\mu\nu}$ and $\phi$, 
\be
S_{\text{mat}}=\int d^4x \sqrt{-\bar g}\sqrt{1-\Gamma\bar X}\left[-\frac12{ \bar g^{\mu\nu}\partial_\mu \sigma \partial_\nu \sigma}-\frac{\Gamma}{2(1-\Gamma\bar X)}\left(\bar g^{\mu\nu}\partial_\mu\sigma\partial_\nu \phi\right)^2 -V(\sigma)\right]\,.
\ee
The equation of motion for $\sigma$ is obtained by varying this action with respect to $\sigma$. Since each field is at most derived once in the action, the equation of motion for $\sigma$ will be second order. The same conclusion holds with the matter contribution to the equation of motion of $\phi$. Therefore,  the presence of a matter scalar field does not introduce higher-order derivative terms in the  equations of motion.

\subsection{Equations of motion} \label{sec-6:5}
Using a disformal transformation,  we provide a new example of naively higher-derivative equations of motion which can be reduced to second-order ones. 
We consider a subclass of G$^3$ theories that can be mapped into Horndeski (where they appear with the metric $\bar g_{\mu \nu}$) and are minimally coupled to matter with their usual metric $g_{\mu \nu}$.
The associated action can thus be written in the form
\be
\label{G3-H}
S=\int d^4 x \sqrt{-\bar g}\, L^H[\bar g_{\mu\nu},\phi]+\int d^4 x \sqrt{-g}\, L_m[g_{\mu\nu}]\,,
\eeq
with
\be
\bar g_{\mu \nu}  =  g_{\mu \nu} + \ga( \phi,  X)\,  \partial_\mu   \phi \, \partial_\nu  \phi \, .
\ee
Since the theory, written in terms of $g_{\mu \nu}$, is not of the Horndeski type, one expects to find higher derivatives in the equations of motion. We show below how to reduce such a system of equations to a second order system.

The variation of the action (\ref{G3-H}) yields
\be
\delta S= \int d^4 x \sqrt{-\bar g}\, \left[{\cal O}_H^{\mu\nu}\delta \bar g_{\mu\nu}+{\cal S}_H \, \delta\phi\right]+\frac12 \int d^4 x \sqrt{-g}\, T_m^{\mu\nu}\delta g_{\mu\nu},
\ee
with 
\be
 \delta \bar g_{\mu\nu}=\delta g_{\mu\nu}+\ga_X \partial_\mu \phi\,  \partial_\nu \phi \delta X+\ga_\phi\,  \partial_\mu \phi \, \partial_\nu \phi \, \delta\phi+2\ga \partial_{(\mu} \phi \nabla_{\nu)}\delta\phi
\ee
and 
\be
\delta X= -\partial^\mu \phi \, \partial^\nu \phi  \delta g_{\mu\nu}+2 \partial^\mu \phi \nabla_\mu\delta\phi\,.
\ee 
The operators ${\cal O}_H^{\mu\nu}$ and ${\cal S}_H$, when expressed in terms of $\bar g_{\mu\nu}$ and $\phi$, contain only second order derivatives since they  come from a Horndeski Lagrangian. 
Variation of the action with respect to the metric $g_{\mu \nu}$ gives the equations of motion 
\be
\label{EOM_g}
{\cal O}_H^{\mu\nu}-{\cal O}_H^{\alpha\beta}\partial_\alpha \phi\, \partial_\beta \phi\,  \ga_X \partial^\mu \phi \, \partial^\nu \phi +\frac12 \Xi \, T_m^{\mu\nu}=0\,,
\ee
where
\be
\Xi\equiv \frac{\sqrt{-g}}{\sqrt{-\bar g}}= \frac{1}{\sqrt{1 + \ga X}}  \;,
\ee
and we used eq.~\eqref{sr} for the second equality.
Variation with respect to $\phi$ gives the scalar equation of motion:
\be \label{scalar-eq}
2\nabla_\mu\left[{\cal O}_H^{\alpha\beta}\partial_\alpha \phi\, \partial_\beta \phi\,  \ga_X \partial^\mu \phi +{\cal O}_H^{\mu\nu} \partial_\nu \phi\,  \ga\right]- {\cal O}_H^{\alpha\beta}\partial_\alpha \phi\, \partial_\beta \phi \, \ga_\phi- {\cal S}_H=0\,.
\ee
Contracting (\ref{EOM_g}) with $\partial_\mu \phi \partial_\nu \phi$ yields
\be
{\cal O}_H^{\alpha\beta}\partial_\alpha \phi\partial_\beta \phi=- \frac{\Xi \, T_m^{\alpha\beta}\partial_\alpha \phi\, \partial_\beta \phi}{2(1-X^2 \ga_X)}\,.
\ee
Substituting back in (\ref{EOM_g}) gives the equation of motion for $g_{\mu\nu}$,
\be
\label{gmunuEOM}
{\cal O}_H^{\mu\nu}=- \frac{\Xi \, T_m^{\alpha\beta}\partial_\alpha\phi\, \partial_\beta \phi}{2(1-X^2 \ga_X)} \ga_X \partial^\mu \phi \, \partial^\nu \phi -\frac12 \Xi \, T_m^{\mu\nu}\,,
\ee
which is second order with respect to $g_{\mu\nu}$. However, it also contains third order derivatives of $\phi$ since ${\cal O}_H^{\mu\nu}$ is second order in $\bar g_{\mu\nu}$, 
which itself depends on the gradient of $\phi$. By taking the trace of~\eqref{gmunuEOM}, one can find a relation expressing the third time derivative of $\phi$ in terms of at most second-order time derivatives.
In this way, the equations of motion~\eqref{gmunuEOM} are effectively second order in time derivatives.
Finally, substituting equation~\eqref{gmunuEOM} in the scalar equation~\eqref{scalar-eq}, one gets
\be
\nabla_\mu\left[\Xi\, \ga_X \frac{(1+\ga X)T^{\alpha\beta}\partial_\alpha \phi\partial_\beta \phi}{1-X^2 \ga_X} \partial^\mu \phi+\Xi \ga T^{\mu\nu} \partial_\nu \phi \right]-\frac12 \Xi\ga_\phi \frac{T^{\alpha\beta}\partial_\alpha \phi\partial_\beta \phi}{1-X^2 \ga_X}+{\cal S}_H=0\,,
\ee 
which is manifestly second order. This  procedure extends that given in~\cite{Zumalacarregui:2013pma} and illustrates how equations of motion that at first view look higher order can in fact be only second order. 

%%%%%%%%%%%%%%%%%%%%%%%%%%%%%%%%%%%%%%%%%%%%%%%%%%%%%%%%%%%%%%%%%%%%%%%%%
%%%%%%%%%%%%%%%%%%%%%%%%%%%%%%%%%%%%%%%%%%%%%%%%%%%%%%%%%%%%%%%%%%%%%%%%%
\section{Conclusions}
%%%%%%%%%%%%%%%%%%%%%%%%%%%%%%%%%%%%%%%%%%%%%%%%%%%%%%%%%%%%%%%%%%%%%%%%%
%%%%%%%%%%%%%%%%%%%%%%%%%%%%%%%%%%%%%%%%%%%%%%%%%%%%%%%%%%%%%%%%%%%%%%%%%

Since its original appearance in~\cite{NRT}, the galileon mechanism has proved an essential tool for modified gravity. Several concrete modified gravity proposals happen to have galileons as their basic skeleton and reduce to galileons in the appropriate decoupling limit. This leads to the possibility of classifying  modified gravity scenarios according to the different inequivalent ways in which the galileons can be consistently coupled to gravity, or ``covariantized". For instance, massive gravity models can be seen as non-minimal covariantizations of the galileon~\cite{deRham:2010ik}, because they involve other degrees of freedom than simply the metric and the scalar field. 
If we insist on  having the minimal number of degrees of freedom and equations of motion strictly of second order in derivatives, we end up in the realm of Horndeski---or generalized galileons, G$^2$ theories
\cite{horndeski,Deffayet:2011gz}.  

In  this paper we have studied in details the scalar-tensor theories proposed in \cite{Gleyzes:2014dya}, called here G$^3$. This class of theories, presented in Sec.~\ref{G3},  covariantizes the galileons in a minimal way, i.e.~without introducing any other degree of freedom than the metric and a scalar field. However, they extend Horndeski in containing two more free functions. They can display equations of motion with derivatives higher than second order in some gauges, but such higher derivatives are in fact harmless, in the sense that they do not bring in unwanted extra degrees of freedom, as we have shown with a detailed Hamiltonian analysis in Sec.~\ref{H}. It turns out that the direct covariantization of the original galileons proposed in~\cite{NRT}, obtained by simply substituting ordinary derivatives with covariant ones, belongs to our class of theories. As such, original galileons are  ``ready to go" without the need of the gravitational counterterms prescribed in~\cite{Deffayet:2009wt}. Contrarily to what was previously thought, their simple minimally coupled versions are free of ghosts instabilities.

Despite the aspects of ``minimality" just discussed, the covariant form of G$^3$ theories is mathematically challenging, due to the high number of derivatives and complexity of the equations involved. We have highlighted  a few ``handles"  to manage their basic properties. First of all, the unitary gauge formulation based on a $3+1$ ADM decomposition, eq.~\eqref{Lagrangian_uni}, is particularly compact and reveals the basic healthy structure of the dynamical system that we are considering. Indeed, the expressions~\eqref{Lagrangian_uni} only contain ``velocities", i.e., \emph{first} time derivatives of the dynamical variables. 

Important insights about scalar-tensor theories can also be given by field transformations. The simplest well-known example is constituted by Brans-Dicke theories, that maintain their basic form under a conformal rescaling of the metric tensor that depends only on the scalar field. On the other hand, the structure of our G$^3$ theories  is invariant under \emph{disformal transformations}, as we discussed in some detail in Sec.~\ref{D}. In particular, disformal transformations with the conformal factor depending on the scalar field only, and the disformal one depending on both $\phi$ and $X$,
\beq \label{morphism}
\tilde g_{\mu\nu}=\Omega^2(\phi) g_{\mu\nu}+\ga(\phi,X) \partial_{\mu}\phi \partial_\nu \phi \,,
\eeq
are the most general class of transformations that preserve the basic G$^3$ structure. This is analogous to the role played by  disformal transformations with $\ga = \ga(\phi)$ for Horndeski theories, which leave them invariant~\cite{DisInv}. We have showed that by applying eq.~\eqref{morphism} to a Horndeski theory we end up in G$^3$---another way of proving the soundness of the corresponding G$^3$ theory---but that, conversely, not all G$^3$ theories can be reduced to the Horndeski form by using~\eqref{morphism}. 

Finally, disformal transformations also help understanding another remarkable property of G$^3$:
even when \emph{minimally} coupled to ordinary matter, G$^3$ exhibit a kinetic type coupling, leading   to a mixing of the dark energy and matter  sound speeds, and thus to a modified Jeans phenomenon~\cite{Gleyzes:2014dya}. In linear perturbations theory, in order to isolate the scalar propagating degree of freedom, one is implicitly de-mixing the scalar from the metric with a field redefinition. For Brans-Dicke theories this can be done at full non-linear level by simply going to the \emph{Einstein-frame} metric with a conformal transformation. In our more general set of theories the mixing terms between the scalar and the metric can be higher in derivatives, in which case they are weighted by the parameter $\alpha_H$, with which we measure the departure from Horndeski. However, 
it is still possible to perform the de-mixing, at least at the \emph{linear} level in perturbation theory. As we show in App.~\ref{app:CC}, part of the field redefinitions~\eqref{toE} that bring us to this generalized Einstein frame corresponds to a disformal transformation of the type discussed above. Because such transformations contain higher derivatives, it ends up mixing matter with the scalar field at a higher order in derivatives, thus affecting the speed of sound of both components. 
The phenomenology of G$^3$, which includes this type of mixing, is an interesting development of this work that we intend to pursue in the future.\footnote{Besides dark energy, the other playground for these theories is inflation, as recently considered in \cite{Fasiello:2014aqa}.}
\\
\\
{\bf Note added}: While finishing this paper, Ref.~\cite{Lin:2014jga} appeared with an analysis and results similar to those of our Sec.~\ref{H}
\\
\\
{\bf Acknowldgements} 
We would like to thank Eugeny Babichev, Christos Charmousis, Claudia De Rham, Kazuya Koyama, Xian Gao, Cristiano Germani, Shinji Mukohyama, David Pirtskhalava, Dani\`ele Steer, Andrew Tolley, Enrico Trincherini, and Miguel Zumalac\'arregui for interesting discussions. D.L. is partly supported by the ANR (Agence Nationale de la Recherche) grant STR-COSMO ANR-09-BLAN-0157-01. J.G.~and F.V.~acknowledge financial support from {\em Programme National de Cosmologie and Galaxies} (PNCG) of CNRS/INSU, France and thank PCCP and APC for kind hospitality. The  work of F.P. is supported by the DOE under contract DE-FG02-11ER41743 and by the A*MIDEX project (n¡ ANR-11-IDEX-0001-02) funded by the ``Investissements dÕAvenir" French Government program, managed by the French National Research Agency (ANR).

\appendix

%%%%%%%%%%%%%%%%%%%%%%%%%%%%%%%%%%%%%%%%%%%%%%%%%%%%%%%%%%%%%%%%%%%%%%%%%
%%%%%%%%%%%%%%%%%%%%%%%%%%%%%%%%%%%%%%%%%%%%%%%%%%%%%%%%%%%%%%%%%%%%%%%%%
\section{Covariant theory}
%%%%%%%%%%%%%%%%%%%%%%%%%%%%%%%%%%%%%%%%%%%%%%%%%%%%%%%%%%%%%%%%%%%%%%%%%
%%%%%%%%%%%%%%%%%%%%%%%%%%%%%%%%%%%%%%%%%%%%%%%%%%%%%%%%%%%%%%%%%%%%%%%%%

%\subsection{From the unitary gauge to the covariant Lagrangians}
\label{Covariant1}

Let us give more details on how to go from the Lagrangians in eq.~\eqref{Lagrangian_uni} to their covariant versions, eqs.~\eqref{L22}--\eqref{L55}.
A crucial relation needed for this calculation is 
\be
K_{\mu \nu}  =-  \frac{\phi_{\mu \nu}}{\sqrt{-X}}+n_\mu \nd_\nu + n_\nu \nd_\mu- \frac{1}{2(-X)} n ^{ \lambda} \nabla_{\lambda} X n_\mu n_\nu \label{phimunu}\, ,\qquad
(\nd_\mu \equiv n^\nu \, \nabla_\nu n_{\mu }) \,,
\ee
which follows from (\ref{defN}) and (\ref{defK}).
As the covariantization of $L_2$ is trivial we start from $L_3$. To rewrite $K$ in terms of scalar field quantities we use the trace of eq.~\eqref{phimunu}, $K=-\left(\Box \phi - {\phi^\lambda \nabla_{\lambda} X }/{2X}\right)/\sqrt{-X}$. Integrating by parts the term proportional to $\nabla_\lambda X$ we obtain eq.~\eqref{L33}.

For $L_4$ we replace the 3-d Ricci curvature $R$ in terms of the 4-d one, ${}^{(4)}\!R$, using the Gauss-Codazzi relation, 
\be
\label{GC1}
{}^{(4)}\! R = R - K^2 + K_{\mu \nu} K^{\mu \nu} + 2 \nabla_\mu (K n^\mu - n^\rho \nabla_\rho n^\mu ) \;,
\ee
after which $L_4$
becomes
\be
\label{step1}
L_4=\Afour{}^{(4)}\! R+(\Bfour + \Afour) (K^2 - K_{\mu \nu}K^{\mu \nu}) -2\Afour\nabla_\mu (K n^\mu - \dot n^\mu )\; .
\ee
Then, using eq.~\eqref{phimunu} and that  $ \nd_\mu =  h_\mu^{ \ \nu} \nabla_{\nu} X/(-2X)$,  it is possible to express the quadratic combination of extrinsic curvatures as
\be
K^2 - K_{\mu \nu}K^{\mu \nu} =-\frac{(\Box \phi)^2-\phi_{\mu\nu}\phi^{\mu\nu}}{X}- \frac{\nabla_{\mu} X (Kn^\mu-\dot n^\mu)}{X}\, .
\ee
After an integration by parts on the last term of eq.~\eqref{step1} we obtain 
\be
\begin{split}
L_4 = & \ \Afour \, {}^{(4)}\!R - \frac{\Afour +\Bfour}{X} \big[(\Box \phi)^2 - \phi_{\mu \nu} \phi^{\mu \nu}\big] \nonumber \\
& + 2 \frac{\Afour+\Bfour - 2 X \Afour_{X}}{X^2}(\phi^{\mu} \phi^{\nu} \phi_{\mu \nu} \Box \phi - \phi^{\mu}  \phi_{\mu \nu} \phi_{\lambda} \phi^{\lambda \nu}) \nonumber \\
& + (C_4 + 2 X C_{4X}) \Box \phi + X C_{4 \phi} \, ,
\end{split}
\ee
where the last line comes from rewriting the term proportional to $\Afour_{\phi}$ analogously to $L_3$ above. 
This equation can be rewritten as eq.~\eqref{L44} by using the definition of $L_4^{\rm gal,1}$ in eq.~\eqref{L44g} and eqs.~\eqref{L2}--\eqref{L4}.

The  case of $L_5$ is the most cumbersome. In addition to the relations \eqref{phimunu} and \eqref{GC1}, we will also need the Gauss Codazzi relation 
\be
\R_{\mu \nu} =  \big({}^{(4)}\!R_{\mu \nu} \big)_\parallel + \big(n^\sigma n^\rho {}^{(4)}\!R_{\mu \sigma \nu \rho} \big)_\parallel - K K_{\mu \nu} + K_{\mu \sigma} K^\sigma_{\ \nu}, \label{GC2}
\ee
where a symbol $\parallel$ denotes the projection on the hypersurface of all tensor indices, {\em e.g.}~$(V_\mu)_\parallel \equiv h^{\ \nu}_{\mu} V_\nu $. For simplicity, let us treat  the two parts of $L_5$ separately. 
Using eq.~\eqref{phimunu} we can rewrite the term proportional to $A_5$ as
\be
\begin{split}
\Bfive & \big(K^3 - 3 K K_{\mu \nu}K^{\mu \nu} + 2  K_{\mu \nu}  K^{\mu \rho} K^\nu_{\ \rho} \big) \\
 = &  -\Bfive(-X)^{-3/2}\big[ (\Box \phi)^3 - 3 (\Box \phi) \phi_{\mu \nu} \phi^{\mu \nu} +2 \phi_{\mu \nu} \phi^{\nu \rho} \phi^\mu_{\ \rho}\big]  \\&+3\Bfive(-X)^{-3/2}
 \bigg[- \frac12 \phi^{\rho} \nabla_\rho X (K^2- K_{\mu \nu}K^{\mu \nu}  ) - 2 (-X)^{3/2}(K \nd_\mu \nd^\mu - K_{\mu \nu}\nd^\mu \nd^\nu) \bigg]\,.
\end{split}
\ee
As we did for $L_3$, we define an auxiliary function, $F_5$,  satisfying $\frac{F_5}{2X}+F_{5X}=\Bfive(-X)^{-3/2}$ and integrate by parts the last line so that
 up to boundary terms the above equation reads,
\be
\label{K3}
\begin{split}
\Bfive & \big(K^3 - 3 K K_{\mu \nu}K^{\mu \nu} + 2  K_{\mu \nu}  K^{\mu \rho} K^\nu_{\ \rho} \big) \\
 = &  -\Bfive(-X)^{-3/2}\big[ (\Box \phi)^3 - 3 (\Box \phi) \phi_{\mu \nu} \phi^{\mu \nu} +2 \phi_{\mu \nu} \phi^{\nu \rho} \phi^\mu_{\ \rho}\big]  \\&- 3 F_{5} \sqrt{-X}\bigg[ \frac12 \big(K^3 - 3 K K_{\mu \nu}K^{\mu \nu} + 2  K_{\mu \nu}  K^{\mu \rho} K^\nu_{\ \rho} \big)  + K^{\mu \nu} n^\sigma n^\rho {}^{(4)}\!R_{\mu \sigma \nu \rho} \\ & -3  K n^\sigma n^\rho  {}^{(4)}\!R_{\sigma \rho} + \nd^\sigma n^\rho {}^{(4)}\!R_{\sigma \rho} \bigg]  + \frac{X}{2} \Ffive_{\phi} (K^2 - K_{\mu \nu} K^{\mu \nu}) \;. \end{split}
\ee

Now we need to deal with the second part. Using the Gauss-Codazzi relations, eqs.~\eqref{GC1} and \eqref{GC2}, this can be rewritten as
\be
\begin{split}
\Afive  \, K_{\mu\nu}\G^{\mu\nu}= & \ \Afive \bigg[K_{\mu\nu} {}^{(4)}\!G^{\mu\nu}+K_{\mu\nu}n_\sigma n_\rho{}^{(4)}\!R^{\mu\sigma\nu\rho}-Kn_\sigma n_\rho{}^{(4)}\!R^{\sigma\rho} \\ & +\frac{1}2 \big(K^3 - 3 K K_{\mu \nu}K^{\mu \nu} + 2  K_{\mu \nu}  K^{\mu \rho} K^\nu_{\ \rho} \big) \bigg]\, .
\end{split}
\ee
We can now replace  $K_{\mu \nu} {}^{(4)}\!G^{\mu\nu}$ using eq.~\eqref{phimunu} and again $ \nd_\mu =  h_\mu^{ \ \nu} \nabla_{\nu} X/(-2X)$. Introducing a new auxiliary function defined as $ G_{5}  \equiv - \int  \Afive_{X}(-X)^{-1/2} \,dX$,
and integrating by parts, first on the $\phi_{\mu\nu}$ term, then on the $\Afive_X$ term that appears,
we finally obtain
\be
\begin{split}
\label{Gmunu}
\Afive  K_{\mu\nu}G^{\mu\nu}=& \ G_5\phi_{\mu\nu} {}^{(4)}\!G^{\mu\nu}+\left(\frac{\Afive_\phi}{\sqrt{-X}}+G_{5\phi}\right)\phi_\mu\phi_\nu {}^{(4)}\! G^{\mu\nu} \\
&+\Afive\bigg[\frac{1}2 \big(K^3 - 3 K K_{\mu \nu}K^{\mu \nu} + 2  K_{\mu \nu}  K^{\mu \rho} K^\nu_{\ \rho} \big) \\ 
&+ K_{\mu\nu}n_\sigma n_\rho{}^{(4)}\!R^{\mu\sigma\nu\rho}-Kn_\sigma n_\rho{}^{(4)}\!R^{\sigma\rho}  +\nd_\mu n_\nu{}^{(4)}\!R^{\mu\nu}\bigg]\;.
\end{split}
\ee
We can now combine the two parts of $L_5$, eqs.~\eqref{K3} and \eqref{Gmunu}, and use the Gauss-Codazzi relation,
\be
n_\mu n_\nu  {}^{(4)}\!G^{\mu\nu}=\frac1{2} \big({\R+K^2-K_{\mu\nu}K^{\mu\nu}} \big)\, ,
\ee 
to rewrite the term $\phi_\mu \phi_\nu  {}^{(4)}\!G^{\mu\nu}$ in eq.~\eqref{Gmunu}.
To simplify this further, we rewrite the combination of Riemann and Ricci that remains employing again eq.~\eqref{K3} which yields
\be
\begin{split}
L_5=& \ G_5\phi_{\mu\nu} {}^{(4)}\!G^{\mu\nu}-\Bfive(-X)^{-3/2}\big[ (\Box \phi)^3 - 3 (\Box \phi) \phi_{\mu \nu} \phi^{\mu \nu} +2 \phi_{\mu \nu} \phi^{\nu \rho} \phi^\mu_{\ \rho}\big]  \\
&+\left(3\Bfive+X\Afive_X\right)\bigg [\frac{1}2 \big( K^3 - 3 K K_{\mu \nu}K^{\mu \nu} + 2  K_{\mu \nu}  K^{\mu \rho} K^\nu_{\ \rho} \big) \\& +(-X)^{-3/2} (\Box \phi)^3 - 3 (\Box \phi) \phi_{\mu \nu} \phi^{\mu \nu} +2 \phi_{\mu \nu} \phi^{\nu \rho} \phi^\mu_{\ \rho})\bigg]\\
&-\frac X2 \left(G_{5\phi}+\frac{\Afive_\phi}{\sqrt{-X}}\right) R- \frac{X}{2}G_{5\phi}(K^2 - K_{\mu \nu} K^{\mu \nu})\, .
\label{L5laststep}
\end{split}
\ee
For the last step, we rewrite the cubic combination of extrinsic curvatures using eq.~\eqref{phimunu} and rewrite the last line analogously to $L_4$, which finally leads to
\be
\begin{split}
L_5 & =G_5 \, {}^{(4)}\!G_{\mu \nu}\phi^{ \mu \nu} - (-X)^{-3/2} \Bfive \big[ (\Box \phi)^3 - 3 (\Box \phi) \phi_{\mu \nu} \phi^{\mu \nu} +2 \phi_{\mu \nu} \phi^{\nu \rho} \phi^\mu_{\ \rho}\big] \nonumber \\
&- \frac{X \Afive_X + 3  \Bfive}{(-X)^{5/2}}\big[ (\Box \phi)^2 \phi_\mu \phi^{\mu \nu} \phi_\nu - 2 \Box \phi \phi_\mu \phi^{\mu \nu} \phi_{\nu \rho} \phi^\rho  - \phi_{\mu \nu} \phi^{\mu \nu} \phi_\rho \phi^{\rho \lambda} \phi_\lambda+ 2 \phi_\mu \phi^{\mu \nu} \phi_{\nu \rho} \phi^{\rho \lambda} \phi_\lambda \big] \nonumber \\
&  + C_5 \, {}^{(4)}\!R - 2 C_{5X} \, \big[ (\Box \phi)^2 - \phi^{ \mu \nu} \phi_{ \mu \nu} \big] + (D_5 + 2 X D_{5X}) \Box \phi + X D_{5 \phi} \, ,
\end{split}
\ee
where, again, the last line comes from applying the method of $L_4$ to the last line of eq~\eqref{L5laststep}. To rewrite this expression as eq.~\eqref{L55} we use the definition of $L_5^{\rm gal,1}$, eq.~\eqref{L55g}, and eqs.~\eqref{L2}--\eqref{L5}.

%%%%%%%%%%%%%%%%%%%%%%%%%%%%%%%%%%%%%%%%%%%%%%%%%%%%%%%%%%%%%%%%%%%%%%%%%
%%%%%%%%%%%%%%%%%%%%%%%%%%%%%%%%%%%%%%%%%%%%%%%%%%%%%%%%%%%%%%%%%%%%%%%%%
\section{Connection to the building blocks of dark energy}
%%%%%%%%%%%%%%%%%%%%%%%%%%%%%%%%%%%%%%%%%%%%%%%%%%%%%%%%%%%%%%%%%%%%%%%%%
%%%%%%%%%%%%%%%%%%%%%%%%%%%%%%%%%%%%%%%%%%%%%%%%%%%%%%%%%%%%%%%%%%%%%%%%%
\label{app:BB}

The dynamics of cosmological perturbations around a FLRW background in the presence of dark energy and modifications of gravity can be systematically studied using the Effective Field Theory of Dark Energy, introduced in Refs.~\cite{Creminelli:2008wc,EFTOr,Bloomfield:2012ff,GLPV,Bloomfield:2013efa} in the case where dark energy can be described by a single scalar degree of freedom. 
In particular, Ref.~\cite{GLPV}  proposed a minimal description of dark energy and modified gravity encompassing all existing models in terms of quadratic Lagrangian operators leading to at most two derivatives in the equations of motion, the so-called  Building Blocks of Dark Energy. In this section we would like to make the connection between these operators, the unitary gauge Lagrangians in eq.~\eqref{Lagrangian_uni} and the parametrisation introduced in  Ref.~\cite{Sawicki}.

As in \cite{GLPV}, let us consider a Lagrangian which is a function of $N$, $K$, $R$, $\SK$ and $\YY$, where $\SK \equiv K_{ij}K^{ij}$ and $ {\cal Y} \equiv K_{ij} R^{ij}$,
i.e., 
\be
\label{full_action_App}
L = L(N,K,\SK,R, {\cal Y}) \;,
\ee
such as  eq.~\eqref{Lagrangian_uni}. 
To isolate linear perturbations, we focus on the quadratic action. 
This can be expanded at second order in the perturbations  around a flat FLRW metric, $ds^2 = - dt^2 + a^2 (t) d \vec x^2$,
using that $\sqrt{-g}=\sqrt{h}N $ and that $\sqrt{h}|_0 = a^3$ on the background.
Then, integrating by parts the term linear in $K$ and using the background equations of motion (the details of these calculations can be found in \cite{GLPV}) the second-order action can be rewritten as
\be
\label{so_action}
\begin{split}
S_{2} \equiv & \int d^4 x \delta_2 \big( \sqrt{-g} L \big) \\
= &\int \! d^4x\frac{M^2(t)}2\bigg\{ \delta_{2} \Big[\sqrt{-g} \Big( {}^{(4)}\!R  -6 H^2   + {2 \rho_m}/{M^2}  - \frac2{N} \big(2 \dot H +  (\rho_m+p_m)/M^2 \big) \Big) \Big] \\ 
&+2 H \alpha_M (t)  \delta_2 \big[ \sqrt{h} (K - 2 H) \big] 
+\alpha_T(t)\, \delta_2 \big( \sqrt{h}  \R  \big)\\
& + a^3 H^2\alpha_K (t) \, \delta N^2 + 4 a^3H\alpha_B(t) \, \delta N \delta K + {a^3}  \alpha_H (t) \, \delta N \R\bigg\}\, ,
\end{split}
\ee
where we have introduced the time-dependent quantities 
\be
\label{Sparameters}
\begin{split}
M^2 & \equiv 2 L_{\SK} \, , \\
\alpha_M & \equiv \frac{\dot L_\SK}{H L_\SK} \;,\\
\alpha_K & \equiv \frac{2L_N +L_{NN}}{2H^2 L_{\SK}}\, , \\  
\alpha_B & \equiv \frac{ 2 H L_{\SK N} + L_{K N}}{4H L_{\SK} }\, , \\
\alpha_T  & \equiv \frac{L_R + \dot L_\YY / 2 + 3 H L_\YY/3}{L_\SK} - 1 \;, \\
\alpha_H & \equiv \frac{L_R + L_{NR} + 3 H L_{\YY}/2 +H L_{N\YY} }{L_\SK} -1 \;,
\end{split}
\ee
evaluated on the background.  Notice that to remove the dependence of  action \eqref{full_action_App} on $\YY$ and obtain eq.~\eqref{so_action} we have used the relation \cite{GLPV}
\be
\lambda(t) \YY = \frac{\lambda(t)}{2} R K + \frac{\dot \lambda(t)}{2N} R \;,
\ee
valid up to boundary terms. 

For a constant $M $, the first line of action \eqref{so_action} describes second-order metric perturbations in a $\Lambda$CDM universe.
The parameters in eq.~\eqref{Sparameters} appear naturally as the coefficients of the second-order expansion of $L$  beyond this standard case. This  expansion makes it also clear that these are the minimal number of parameters describing the dynamics once the 
background  expansion history, $H(t)$, and the matter content, i.e.~$\rho_m (t_0)$ and its equation of state, are given.

Not surprisingly, the first 5 of these parameters are the same as those proposed in Ref.~\cite{Sawicki}. 
The last one is new and parameterizes a deviation from Horndeski theories. Using $L=L_2+L_3+L_4+L_5$, in eq.~\eqref{ABparameters} we have
written these parameters in terms of the functions $A_i$ and $B_i$ appearing in the Lagrangians \eqref{Lagrangian_uni}.

In Ref.~\cite{GLPV} we explicitly separated the operators affecting the perturbations from those fixed by the background evolution, writing the action as 
\be
\begin{split}
\label{EFT_action}
S= \int \! d^4x \sqrt{-g} \bigg[ \, &\frac{\MM^2}{2} f(t) {}^{(4)}\!R - \Lambda(t) - c(t) g^{00}  + \, \frac{M_2^4(t)}{2} (\delta g^{00})^2\, -\, \frac{m_3^3(t)}{2} \, \delta K \delta g^{00} \, 
  \\[1.2mm]
 &  -  \,  m_4^2(t)\left(\delta K^2 - \d K^\mu_{ \ \nu} \, \d K^\nu_{ \ \mu} \right) \, +\, \frac{\tilde m_4^2(t)}{2} \, \R \, \delta g^{00}  \bigg] \;.
\end{split}
\ee
As explained in \cite{EFTOr,Bloomfield:2012ff,GLPV,Bloomfield:2013efa}, the functions $c$ and $\Lambda$ are fully specified by the background expansion history.
We are thus left with $6$ free parameters in this action. As expected, there is a simple relation between these parameters and those in eq.~\eqref{Sparameters}.
Indeed, at second order the above action reduces to eq.~\eqref{so_action} with the following dictionary between the two notations,
\be
\begin{split}
M^2 & =\MM^2f+2\mfs \, , \\
\alpha_M & =\frac{2 \dot M}{MH}\, , \\
\alpha_K & = \frac{2c+4M_2^4}{M^2H^2}\, , \\  
\alpha_B & =\frac{\MM^2 \dot f - m_3^3}{2 M^2H}\, , \\
\alpha_T  & =-\frac{2\mfs}{M^2} \;, \\
\alpha_H &=\frac{2(\tmfs-\mfs)}{M^2} \, .
\end{split}
\ee
To see this, one can  use $g^{00}=-1/N^2$ and rewrite the term proportional to $c$ up to second order as
\be
- c g^{00} = - \frac{c}{N} (1- \delta N) -  c \, \delta N^2 \;.
\ee 
The last term combines with the operator proportional to $M_2^4$.
Moreover, one can rewrite the term proportional to $m_4^2$, up to boundary terms, as 
\be 
\begin{split}
m_4^2 \left(\delta K^2 - \d K^\mu_{ \ \nu} \, \d K^\nu_{ \ \mu} \right) &= m_4^2 \Big( {}^{(4)}\! R - R - 6H^2 + 4H K \Big) + 2 (m_4^2)^{\hbox{$\cdot$}} \frac{K}{N} \\
& = m_4^2 \Big( {}^{(4)}\! R - R \Big) + \big[ \MM^2 \dot f + 2 (m_4^2)^{\hbox{$\cdot$}} \big] \frac{K}N + \MM^2 \dot f \delta N \delta K  + \frac{\MM^2 \ddot f}{N} + 3 H \MM^2 \dot f \frac{\delta N}{N} \;,
\end{split}
\ee
and use the  background equations of motion for the last two terms.

Finally, it is also easy to make connection with the (slightly different) notation adopted in~\cite{Piazza:2013pua}, where the phenomenological aspects of dark energy were studied by using the formalism developed in~\cite{EFTOr,Bloomfield:2012ff,GLPV,Bloomfield:2013efa,PV}.
There, the time-dependent ``Planck mass squared" $\MM^2 f(t)$ was pulled out of the action,
\be \label{example}
\begin{split}
S \ =\  &  \int \! d^4x \, \sqrt{-g} \, \frac{\MM^2 f(t)}{2} \, \Big[{}^{(4)}\!R \, -\,  2 \lambda(t) \, - \, 2 {\cal C}(t) g^{00}   \Big. \\[1.2mm]
& \left.+ \, \mu_2^2(t) (\delta g^{00})^2\, -\, \mu_3(t) \, \delta K \delta g^{00} 
   + \,  \epsilon_4(t) \left(\delta K^\mu_{ \ \nu} \, \delta K^\nu_{ \ \mu}  - \delta K^2  \right) +  \frac{\tilde \epsilon_4(t)}{2}  R\,   \delta g^{00} 
  \right] \;, \\
\end{split}
\ee
so that the natural order of magnitude of the time-dependent coefficients (inside the square brackets above) is the Hubble parameter to the appropriate power. This is also evident by the following dictionary
\be
\begin{split}
M^2 & =\MM^2f\, (1+ \epsilon_4)\, , \\
\alpha_M & =\frac{\dot \epsilon_4}{H(1+ \epsilon_4)} + \frac{\mu}{H}\, , \\
\alpha_K & = \frac{2{\cal C}+4\mu_2^2}{H^2 (1+ \epsilon_4)}\, , \\  
\alpha_B & =\frac{\mu - \mu_3}{2 H (1+ \epsilon_4)}\, , \\
\alpha_T  & =-\frac{\epsilon_4}{1+\epsilon_4} \;, \\
\alpha_H &=\frac{\tilde \epsilon_4 - \epsilon_4}{1+\epsilon_4} \, .
\end{split}
\ee
%%%%%%%%%%%%%%%%%%%%%%%%%%%%%%%%%%%%%%%%%%%%%%%%%%%%%%%%%%%%%%%%%%%%%%%%%
%%%%%%%%%%%%%%%%%%%%%%%%%%%%%%%%%%%%%%%%%%%%%%%%%%%%%%%%%%%%%%%%%%%%%%%%%
\section{Disformal transformation in Newtonian gauge}
%%%%%%%%%%%%%%%%%%%%%%%%%%%%%%%%%%%%%%%%%%%%%%%%%%%%%%%%%%%%%%%%%%%%%%%%%
%%%%%%%%%%%%%%%%%%%%%%%%%%%%%%%%%%%%%%%%%%%%%%%%%%%%%%%%%%%%%%%%%%%%%%%%%
\label{app:CC}

In this appendix we show that (part of) the change of variables introduced in Sec.~\ref{HEsect} in order to de-mix the metric Newtonian potentials and the scalar field can be understood in terms of a disformal transformation.
In particular, we restrict to the G$^3$ Lagrangian $L_4$ in eq.~\eqref{Lagrangian_uni}, given 
in terms of the metric $g_{\mu \nu}$, and 
we show that after the disformal transformation \eqref{disformal_L4} with $\ga = \ga_4$ satisfying eq.~\eqref{Lambda_4X}, all the couplings proportional to $\alpha_H$ disappear from action \eqref{KinetAct}. To maintain the usual background time-time component of the barred metric, $\bar g_{00}^{(0)}=-1$, together with the field redefinition \eqref{disformal_L4} we also perform a time coordinate change,
\be
\label{time}
\bar t = \int \sqrt{1-\ga_0} \, dt -\alpha \, , \qquad \alpha \equiv \frac{\ga_0}{\sqrt{1-\ga_0}} \pi\;,
\ee
where $\ga_0$ is the background value of $\ga$.
The change $t \to t -\alpha$ ensures that $\bar g_{0i} = g_{0i}$ and that we  thus remain in Newtonian gauge (see eq.~\eqref{NG}).
Using $\phi = t + \pi$, the combination of eq.~\eqref{disformal_L4} and the above time redefinition gives, up to linear order,
\be
\bar g_{00} = \frac{g_{00} + \ga (1+2 \dot \pi)}{ 1 -\ga_0} - 2 \frac{d\alpha}{d\bar t} \;, \qquad \bar g_{0i} = g_{0i} =0 \;, \qquad \bar g_{ij} = g_{ij}\;,
\ee
where a dot always denotes the derivative with respect to $t$. Expanding $\ga$ to linear order and defining $\bar g_{00} \equiv -(1+2 \bar \Phi)$ and $\bar g_{ij} =  a^2 (\bar t) (1-2 \bar \Psi) \delta_{ij}$, 
we obtain, for the potentials in the barred frame,
\be 
\bar \Phi = \frac{ (1 - \ga_{X})\Phi + \ga_{X} \dot \pi }{1-\ga_0} + \frac{\dot \ga_0 \pi }{2 (1- \ga_0)^2} \;, \qquad \bar \Psi = \Psi - \frac{\ga_0}{1-\ga_0} H \pi\;.
\ee
Since the time has been redefined according to eq.~\eqref{time}, $\pi$ in the barred frame reads
\be
\bar \pi = \frac{1}{\sqrt{1-\ga_0}} \pi\;,
\ee
where we have used $ \pi = - \delta t$ and $\bar \pi = - \delta \bar t$.

We can rewrite the time dependent quantities $\ga_0$, $\dot \ga_0$ and $\ga_X$ in terms of the quantities $\alpha_i$ and $\bar \alpha_i$, using the definitions of $\alpha_i$ in eq.~\eqref{ABparameters} together with the metric transformation \eqref{disformal_L4} and eqs.~\eqref{barA4toA4} and \eqref{barB4toB4}. This yields
\be
1-\ga_0 = \frac{1+\alpha_T}{1+ \bar \alpha_T} \;, \qquad \dot \ga_0 = \frac{1+\alpha_T}{1+ \bar \alpha_T} (\alpha_M - \bar \alpha_M)\;, \qquad \ga_{X} = - \alpha_H \;.
\ee
Replacing these relations in the above equations and using $\bar H =  H/\sqrt{1 - \ga_0}$ due to the time redefinition, we obtain
\be
\begin{split}
\bar \Phi &= \frac{1 + \bar \alpha_T}{1 + \alpha_T} \left[ (1+ \alpha_H) \Phi + (\alpha_M - \bar \alpha_M) H \pi - \alpha_H \dot \pi  \right] \;, \\
\bar \Psi &= \Psi + \frac{\alpha_T - \bar \alpha_T}{1 + \alpha_T} H \pi \;,  \\
\bar \pi &= \frac{1 + \bar \alpha_T}{1 + \alpha_T}  \frac{H}{\bar H } \pi \;.
\end{split}
\ee 
These are the field redefinitions in Netwonian gauge between the two frames. One can use these relations, together with an expression for $\bar \alpha_B$ and $\bar \alpha_K$ as a function of the other quantities,
to rewrite action \eqref{KinetAct} in  the barred frame, where all the couplings proportional to $\alpha_H$ disappear. 
Here we simply check, using the relations above and 
\be
{\bar \alpha_B}=-1 + \frac{1 + \alpha_B}{1+\alpha_H} \frac{1+ \alpha_T}{1+ \bar \alpha_T} \, , 
\ee
that $\Phi_E$ and $\Psi_E$ given in eq.~\eqref{toE} become, as expected,
\be
\begin{split}
 \Phi_E & =    \frac{1}{1+ \bar \alpha_T} \bar \Phi +   \bigg( \frac{1+\bar \alpha_M}{1+\bar \alpha_T} - 1 -  \bar \alpha_B \bigg) \bar H \bar \pi   \; , \\
 \Psi_E & = \bar \Psi - \bar \alpha_B   \bar H \bar \pi \, .
\end{split}
\ee

%%%%%%%%%%%%%%%%%%%%%%%%%%%%%%%%%%%%%%%%%%%%%%%%%%%%%%%%%%%%%%%%%%%%%%%%%
%%%%%%%%%%%%%%%%%%%%%%%%%%%%%%%%%%%%%%%%%%%%%%%%%%%%%%%%%%%%%%%%%%%%%%%%%

\end{document}